\documentclass[12pt,preprint]{aastex}
\usepackage{apjfonts}
\usepackage{epsfig}
\usepackage{amsmath}
\begin{document}

\title{Statistical study of gamma-ray bursts with jet break feature in multi-wavelength afterglow emissions}
\author{Wen Zhao$^{1}$, Jia-Chang Zhang$^{1}$, Qing-Xiang Zhang$^{1}$, Jian-Tong Liang$^{1}$, Xiao-Hang Luan$^{1}$, Qi-Qi Zhou$^{1}$, Shuang-Xi Yi$^{1}$, Fei-Fei Wang$^{2}$ and Shao-Tong Zhang$^{2}$}
\affil{$^{1}$School of Physics and Physical Engineering, Qufu Normal University, Qufu 273165, China; yisx2015@qfnu.edu.cn\\
       $^{2}$School of Mathematics and Physics, Qingdao University of Science and Technology, Qingdao 266061, China}

\begin{abstract}
It is generally supposed that a transition from the normal decay phase (decay slope $\sim -1$) to a steeper phase (decay
slope $\sim -2$) could be suggested as a jet break. The jet opening angle $\theta_{\rm jet}$ is then calculated from the jet break time of the afterglow light curve. This allows the derivation of the collimation-corrected energy $E_{\rm jet}$ of those GRBs. We extensively searched for the GRBs with jet break features from multi-wavelength afterglow light curves, and 138 GRBs with significant breaks were collected. The jet break times of those GRBs mainly range from 1000 s to $10^6$ s, and the distribution of the collimation-corrected energy $E_{\rm jet}$ peaks at $\sim10^{50}$ erg. We also confirmed the $E_{\rm \gamma,iso}-E_{\rm p,i}$, $E_{\rm jet}-E_{\rm p,i}$ and $E_{\rm \gamma,iso}-\theta_{\rm jet}$ relations, and found $E_{\rm \gamma,iso}-T_{\rm j,z}-E_{\rm p,i}$ relation remains tight with more multi-wavelength data. This tight $E_{\rm \gamma,iso}-T_{\rm j,z}-E_{\rm p,i}$ relation is also conformed by different groups of our selected GRBs in the paper. In addition,
another two new and tighter correlations among $E_{\rm jet}-T_{\rm j,z}-E_{\rm p,i}$ are well confirmed for different circumburst mediums in this paper. We suggest that those tight three-parameter correlations are more physical, and could be widely applied to constrain the cosmological parameters.
\end{abstract}


\section{Introduction}
Gamma-ray bursts (GRBs) are supposed to be the most powerful electromagnetic explosions in the universe.
These erratic, luminous, transient events are very different from the other astrophysical phenomena.
Generally speaking, according to the fireball model, GRBs are believed to be produced in a relativistic outflow by the internal shock with different duration times in gamma-ray emission (Piran 2004; M{\'e}sz{\'a}ros 2006; Zhang 2007; Kumar \& Zhang 2015),  therefore, short GRBs ($T_{90} < 2 \,s$) and long GRBs ($T_{90} > 2 \,s$) are proposed in many works. Long GRBs are produced via the core collapse of a massive star (Woosley 1993; MacFadyen \& Woosley 1999), while short
GRBs are thought to be the merger of two compact stars (Paczynski 1986; Eichler et al. 1989; Narayan et al. 1992).
More interestingly, the binary neutron star merger GW 1708171 accompanied by short GRB 170817A was also firstly detected by Advanced LIGO and Advanced VIRGO (Abbott et al. 2017).

The broadband afterglows of GRBs are usually interpreted as the interaction of the fireball shell with the ambient medium (M{\'e}sz{\'a}ros \& Rees 1997; Sari et al. 1998; Zou et al. 2005; Yi et al. 2013; Gao et al. 2013). Different radiation characteristics appeared in the multi-wavelength light curves, such as, the erratic flares and several power-law phases in X-ray emissions (Zhang et al. 2006; Nousek et al. 2006; Liang et al. 2007; Wang \& Dai 2013; Lei et al. 2013; Yi et al. 2016; Liu et al. 2017), the reverse shock features (or flares) and onset bumps in optical afterglows (Kobayashi 2000; Zhang et al. 2003; Japelj et al. 2014; Gao et al. 2015; Liang et al. 2010, 2013; Huang et al. 2016; Yi et al. 2017a, b; Zhou et al. 2020; Yi et al. 2020).
After decades of observations, many multi-wavelength afterglow emissions were obtained, and those different emission components have greatly improved our understanding on the physical origin of GRBs (see Wang et al. 2020, a comprehensive statistical study on GRBs).

However, some temporal break features observed in afterglows indicate that GRBs should be collimated into the narrow
jets (Rhoads 1997; Li et al. 2020). The collimation of GRB jets is very important to study the physics of GRBs, especially for their central engine and true gamma-ray energy emitted from the source. This allows the derivation of the collimation-corrected energy of the GRBs, and has greatly improved our understanding on GRBs. Actually, many papers about the jet break features have been carried out. In the pre-Swift era, some GRBs with jet breaks in the optical band have been observed after several days since the GRB
trigger (Rhoads 1999; Sari et al. 1999; Halpern et al. 2000; Bloom et al. 2001; Frail et al. 2001; Jaunsen et al. 2001; Wei \& Lu 2002; Wu et al. 2004). Even more investigations were applied to research and study the statistical
features of jet breaks based on the XRT data in the Swift era (Burrows et al. 2006; Burrows \& Racusin 2006;
Grupe et al. 2006; Dai et al. 2007; Jin et al. 2007; Nava et al. 2007; Panaitescu 2007; Willingale et al.
2007; Kocevski \& Butler 2008; Liang et al. 2008; de Pasquale et al. 2009; Evans et al. 2009; Racusin
et al. 2009; Wang et al. 2015a).

The study of jet break is very important to explain the physics of GRBs. Liang et al. (2008) explored whether the observed breaks in the afterglow light curves are jet breaks or not, in order to infer GRB energetics. Wang et al. (2018a) revisited the jet breaks with 99 GRBs, and found the temporal and spectral behaviors of 55 GRBs is consistent with the theoretical predictions of the jet break models. Sharma et al. (2020) study the X-ray afterglow jet break properties of GRB 160325A, and suggest that the kinetic energy of the jet is likely dissipated via internal shocks, which evolves from an optically thick to optically thin environment within the jet.

In this paper, we continued to search for the jet break features from multi-wavelength afterglow light curves, investigate the properties and some empirical relations about jet breaks. This paper is organized as follows. The jet break sample selection criteria is presented in Section 2. The parameter characteristics and their correlations are shown in Section 3. Discussion and Conclusions are presented in Section 4.

\section{Sample selection}
Some temporal break features observed in multi-wavelength afterglows indicate that the GRB outflows are
collimated. In order to recognize jet breaks in GRBs, one should systematically identify the potential temporal breaks in the multi-wavelength afterglows. A shallow decay "plateau" segment is commonly appeared in the X-ray and optical afterglow light curves, which generally shows a shallow decay ( $\sim$ -0.5) phase to a normal decay ( $\sim$ -1) segment (Zhang et al. 2006; Nousek et al. 2006; Si et al. 2018). The shallow decay segments are studied frequently, which are interpreted as the energy injection from the central engine of GRB itself (Dai \& Lu 1998a, b; Zhang \& M{\'e}sz{\'a}ros 2001; Dall'Osso et al. 2011; Rowlinson et al. 2013, 2014; L\"u \& Zhang 2014; Rea et al. 2015; Si et al. 2018; Stratta et al. 2018). Some tight correlations about the plateaus have also been obtained, such as, the relation of $L_{\rm x}-T_{\rm b}$ for the X-ray plateaus, which has been used as cosmological tool in Cardone et al. (2009), Cardone et al. (2010), Dainotti et al. (2013) and Postnikov et al. (2014). Another break usually has a transition from the normal decay phase with the decay slope about -1 to a steeper phase segment ($\sim$ - 2), which can be interpreted as a jet break (Rhoads 1999; Sari et al. 1999; Frail et al. 2001; Wu et al. 2004; Liang et al. 2008; Racusin et al. 2009; Wang et al. 2018a).

In our paper, we extensively search the jet
break features from multi-wavelength afterglow light curves that have a transition from the normal decay segment ($\sim$ - 1) to a steeper phase ($\sim$ -2). 138 GRBs with such jet breaks and redshifts are identified in our sample, including two short GRBs 051221A and 090426. Most of the jet breaks are obtained by X-ray afterglow light curves, but some of them are calculated from optical and radio light curves, respectively. The relevant information (such as, redshift z, the isotropic energy $E_{\rm \gamma, iso}$, the peak energy of GRB spectrum in the rest frame $E_{\rm p,i}$, the jet break time $T_{\rm jet}$, frequency band and the references) is collected in Table 1.  A majority of jet break times, including the isotropic energy and peak energy of GRB spectrum, are collected from the published papers or GCNs. But some jet breaks of recent GRBs are identified by X-ray afterglow light curves fitted with an empirical smooth broken power-law function of ourselves (SBPL, Si et al. 2018).
The half-opening angle of GRBs with redshift can be obtained by
the isotropic energy $E_{\rm \gamma, iso}$ and jet break time $T_{\rm jet}$ in
a homogeneous interstellar medium (ISM) case (Rhoads 1999; Sari et al.
1999; Frail et al. 2001; Yi et al. 2015),
\begin{equation}
\theta_{\rm jet}(ISM)=0.076 \,\,{\rm rad}\,\,\left(\frac{T_{\rm jet}}{1\ \rm
day}\right)^{3/8}\left(\frac{1+z}{2}\right)^{-3/8}
E_{\gamma,iso,53}^{-1/8}\left(\frac{\eta}{0.2}\right)^{1/8}\left(\frac{n}{1\ \rm
cm^{-3}}\right)^{1/8},
\end{equation}
or in the wind profile (Chevalier \& Li 2000; Bloom et al. 2003; Yi et al. 2015),
\begin{equation}
\theta_{\rm jet}(Wind)=0.12\,\,{\rm rad}\,\,\left(\frac{T_{\rm jet}}{1\ \rm
day}\right)^{1/4}\left(\frac{1+z}{2}\right)^{-1/4}
E_{\gamma,iso,52}^{-1/4}\left(\frac{\eta}{0.2}\right)^{1/4}A_{*}^{1/4}.
\end{equation}
Where the efficiency of GRBs $\eta=0.2$, the ambient medium $n=1 $ cm$^{-3}$ and the wind
parameter $A_* = 1$ are adopted. Deriving the half opening angles with the isotropic energy
and jet break time for different ambient mediums allows a measurement of the true energetics
of GRBs, i.e., $E_{\rm jet}=E_{\rm \gamma, iso}(1-cos\theta_{\rm jet})$. The jet opening angle
and beaming corrected energy for each GRB of the sample are also presented in Table 1.

\section{Correlations about GRBs with jet breaks}
The following statistics of the parameters applies to the jet break of GRBs. Figure 1 shows the distributions of the selected
GRB parameters, i.e., the isotropic energy $E_{\rm \gamma, iso}$ and jet break time $T_{\rm jet}$, also including the derived jet opening angle
$\theta_{\rm jet}$ and beaming corrected energy $E_{\rm jet}$ for different ISM and wind circumburst mediums. The jet break times of GRBs range from 1000 s to $10^6$ s, with the typical break time $\sim$ a few times of $ 10^4$ s. However, the jet break times of some GRBs occur at late time about $10^6$ s. Most of the selected GRBs have an opening angle $\theta_{\rm jet}$ of $2.5^{\circ}$ for both ISM and wind profile,
and the distribution of the beaming corrected energy $E_{\rm jet}$ of GRBs peaks at $\sim10^{50}$ erg.

Some tight GRB luminosity correlations about light curve breaks were proposed before, such as, $E_{\rm \gamma,iso}-T_{\rm j,z}-E_{\rm p,i}$ (called Liang-Zhang relation, Liang \& Zhang 2005), $L_{\rm b,z}-T_{\rm b,z}$ and $L_{\rm b,z}-L_{\rm peak}$ (Dainotti correlation in two dimension, Dainotti et al. 2008, 2011, 2015), $L_{\rm b,z}-T_{\rm b,z}-L_{\rm peak}$ (Dainotti correlation in three dimension, Dainotti et al. 2016, 2017a), $L_{\rm b,z}-T_{\rm b,z}-E_{\rm \gamma,iso}$ (Xu - Huang relation, Xu \& Huang 2012; Dainotti et al. 2013; Si et al. 2018; Tang et al. 2019), and $L_{\rm b,z}-T_{\rm b,z}-E_{\rm p,i}$ (Si 2018 relation, Si et al. 2018), where $T_{\rm j,z}=T_{\rm j}/(1+z)$, $L_{\rm peak}$ is the burst peak isotropic luminosity, $L_{\rm b,z}$ and $T_{\rm b,z}$  are the plateau break luminosity and time in the rest frame, repectively.
And some others relations for $E_{\rm \gamma, iso}-E_{\rm p,i}$ (Amati relation, Amati et al. 2002, 2008), $E_{\rm jet}-E_{\rm p,i}$ (Ghirlanda relation; Ghirlanda et al. 2004), $L_{\rm p}-E_{\rm p,i}$ (Yonetoku relation, Yonetoku et al. 2004) and $E_{\rm \gamma, iso}-\theta_{\rm jet}$ (Frail relation; Frail et al. 2001). Those correlations are widely used to perform the cosmological parameters (Dai et al. 2004; Wei et al. 2013; Wang et al. 2011; Wang et al. 2015b; Wang et al. 2016; Dainotti \& Del Vecchio 2017; Dainotti \& Amati 2018; Wang \& Wang 2019). For a more comprehensive study about GRB correlations, please see the recent reviews (Wang et al. 2015b; Dainotti et al. 2018; Dainotti 2019 for a book about GRB correlations).
Therefore, it would be meaningful to continue investigating several empirical correlations of GRBs proposed in previous studies with more selected data, we then compile the related parameters of GRBs in our jet break sample in Table 1. Our statistical method is linear regression and Spearman correlation coefficient (Spearman 1987). We present three two-parameter correlations, such as, $E_{\rm \gamma, iso}-E_{\rm p,i}$, $E_{\rm jet}-E_{\rm p,i}$ and $E_{\rm \gamma, iso}-\theta_{\rm jet}$. The best fit results are shown in Table 2. We confirm the tight $E_{\rm \gamma, iso}-E_{\rm p,i}$ correlation, the best linear fitting result is $E_{\rm \gamma, iso} \propto E_{\rm p,i}^{1.27\pm0.04}$ with a Spearman correlation coefficient $R=0.74$ and chance probability $p=4.83\times10^{-24}$. We provide the beaming corrected energy $E_{\rm jet}$ for ISM and wind case in Table 1, therefore we plot $E_{\rm jet}$ versus $E_{\rm p,i}$ for two different ambient mediums, too. We find our results are generally consistent with $E_{\rm jet}-E_{\rm p,i}$ relation, even though the two correlations have large scatter. The isotropic energy $E_{\rm \gamma, iso}$ is anti-correlated with the jet opening angle $\theta_{\rm jet}$, but the correlation for wind case is much tighter compared with the correlation of ISM case. We give the scatter plots between two parameters in Figure 2.

We next evaluate the three-parameter correlation about $E_{\rm \gamma,iso}-T_{\rm j,z}-E_{\rm p,i}$, which is also called Liang-Zhang relation (Liang \& Zhang 2005). This is a tight correlation as proposed by Liang \& Zhang (2005). The previous result is $E_{\rm \gamma,iso} = (0.85 \pm 0.21) \times (E_{\rm p,i})^{1.94 \pm 0.17} \times (T_{\rm j,z})^{-1.24 \pm 0.23}$. Limiting to the optical break sample, they selected 15 GRBs with measurements of the redshift, optical break time and the spectral peak energy in the rest frame. According to Wang et al. (2018a), who selected 55 GRBs with optical jet break features, and they found the $E_{\rm \gamma,iso}-T_{\rm j,z}-E_{\rm p,i}$ relation remains tight. However, Wang et al. (2018a) found this relation is less tight than before, it is mostly likely the early jet breaks and hence small opening angle jets. In our paper, 138 GRBs with jet break features are identified by multi-wavelength afterglow light curves, we continue to explore $E_{\rm \gamma,iso}-T_{\rm j,z}-E_{\rm p,i}$ relation with our multi-wavelength break data. For clarity, we separated our multi-wavelength jet break sample to the optical and X-ray groups, and also presented the $E_{\rm \gamma,iso}-T_{\rm j,z}-E_{\rm p,i}$ relation with the total sample. We found the divided optical and X-ray sample are consistent with the $E_{\rm \gamma,iso}-T_{\rm j,z}-E_{\rm p,i}$ relation well, the Spearman correlation coefficient are $R=0.80$ and $R=0.84$ for the optical and X-ray sample, respectively. The best-fit correlation for the optical sample is shown
\begin{equation}
\log E_{\rm \gamma,iso}=(49.89\pm 0.46)+(0.12\pm0.02)\times \log T_{\rm j,z}+(0.99\pm0.17)\times \log E_{\rm p,i},
\end{equation}
and the $E_{\rm \gamma,iso}-T_{\rm j,z}-E_{\rm p,i}$ relation for X-ray sample in the form of
\begin{equation}
\log E_{\rm \gamma,iso}=(50.22\pm 0.12)-(0.14\pm0.01)\times \log T_{\rm j,z}+(1.23\pm0.04)\times \log E_{\rm p,i}.
\end{equation}
We then considered the $E_{\rm \gamma,iso}-T_{\rm j,z}-E_{\rm p,i}$ relation with the total sample. Including the early and late time jet breaks of our multi-wavelength data, the correlation is still tight with the correlation coefficient is $R=0.80$ and chance probability $p=5.22\times10^{-32}$. The scatter plot is shown in Figure 3, and the best fit results are presented in Table 2. As we can see, our result is also less tight than Liang \& Zhang (2005), maybe it is due to the early jet breaks or the multi-wavelength afterglows. This indicates that the tight correlation of $E_{\rm \gamma,iso}-T_{\rm j,z}-E_{\rm p,i}$ may also exist in the multi-wavelength afterglow light curves. The best-fit correlation is taken as
\begin{equation}
\log E_{\rm \gamma,iso}=(50.05\pm 0.26)-(0.02\pm0.01)\times \log T_{\rm j,z}+(1.13\pm0.09)\times \log E_{\rm p,i}.
\end{equation}
We continued to explore the possible correlations among $E_{\rm jet}-T_{\rm j,z}-E_{\rm p,i}$, we applied the derived parameter $E_{\rm jet}$ for ISM and wind profile to replace the isotropic one in $E_{\rm \gamma, iso}-T_{\rm j,z}-E_{\rm p,i}$, and found there are also two new tight correlations between them for ISM and wind profile, respectively,

\begin{equation}
\log E_{\rm jet}(ISM)=(45.08\pm 0.21)+(0.67\pm0.02)\times \log T_{\rm j,z}+(0.84\pm0.07)\times \log E_{\rm p,i},
\end{equation}
and
\begin{equation}
\log E_{\rm jet}(Wind)=(46.77\pm 0.14)+(0.45\pm0.01)\times \log T_{\rm j,z}+(0.56\dot{}\pm0.05)\times \log E_{\rm p,i}.
\end{equation}
The two new three-parameter correlations are even more tighter than the other correlations discussed in the paper as shown in Figure 4, and both of them have the correlation coefficient $R=0.90$. We suppose those tight three-parameter correlations are more physical, especially for the corrected parameters. These correlations could be used as standard candles, and widely applied to perform the cosmological parameters.

However, as is well known, the observational parameters of GRBs are widely suffered by instrumental selection effects or biases. Some of derivation parameters about the correlations are mainly based on the observed flux, redshift, and temporal variability. Therefore, it should be mentioned that all these correlations appeared in the paper are subjected to selection biases and redshift evolution (Efron \& Petrosian 1992, 1995; Lloyd \& Petrosian 1999, 2000; Dainotti et al. 2013, 2017b). Therefore, these tight two or three parameter correlations in this paper may be, at least partially, suffered by sample selection biases and redshift evolution.

\section{Conclusions and Discussion}
The collimation of GRB fireball is a significant subject, those investigations in this topic could provide us some important clues on the GRB central engines, and also including some empirical correlations of GRBs. The observational phenomena indicate that this scenario has the achromatic break in the multi-wavelength afterglow light curves. Therefore, we extensively search for the multi-wavelength afterglow light curves that have a transition from the normal decay segment ($\sim$ - 1) to a steeper phase ($\sim$ -2), and 138 GRBs with the achromatic break features are collected. Most of the jet breaks are obtained by X-ray afterglow light curves, but some of them are calculated from optical and radio light curves.

The jet break time of selected GRBs has a distribution between 1000 s and $10^6$ s, with the typical break time $\sim$ a few times of $ 10^4$ s. Most of GRBs have a half-opening-angle $\theta_{\rm jet}$ of $2.5^{\circ}$ when considering both ISM and wind profile,
and the distribution of collimation-corrected energy $E_{\rm jet}$ peaks at $\sim10^{50}$ erg. We also investigated several previously claimed empirical correlations with our multi-wavelength data, we found the correlations of $E_{\rm \gamma,iso}-T_{\rm j,z}-E_{\rm p,i}$ (Liang-Zhang) and $E_{\rm \gamma,iso}-E_{\rm p,i}$ (Amati) remain tight, and the tight $E_{\rm \gamma,iso}-T_{\rm j,z}-E_{\rm p,i}$ relation is even conformed by different groups of our selected sample. $E_{\rm jet}-E_{\rm p,i}$ (Ghirlanda) and $E_{\rm \gamma,iso}-\theta_{\rm jet}$ (Frail) relations are still existed for different ISM and wind cases with the selected data, even though those correlations have large scatter.
We continued to provide a multiple linear regression analysis for the collimation-corrected energy $E_{\rm jet}$ and $T_{\rm j,z}-E_{\rm p,i}$.
The analysis of two new correlations among these observables are well confirmed for different circumburst mediums. The two new three-parameter correlations are even more tighter, and both of them have the correlation coefficient $R=0.90$. Those tight empirical relations may directly indicate the radiation physics of GRB outflows.

\section*{Acknowledgments}
We thank the anonymous referee for constructive suggestions.
We also thank Fa-Yin Wang, Yuan-Chuan Zou and Tong Liu for helpful discussion.
This work is supported by the National Natural Science Foundation of China (Grant No. 11703015),
the Natural Science Foundation of Shandong Province (Grant No. ZR2017BA006), the "provincial university student innovation and entrepreneurship training programs" of Qufu Normal University (S201910446012), and the Youth Innovations and Talents Project of Shandong Provincial Colleges and Universities (Grant No. 201909118).

\begin{figure*}
\includegraphics[angle=0,scale=0.35]{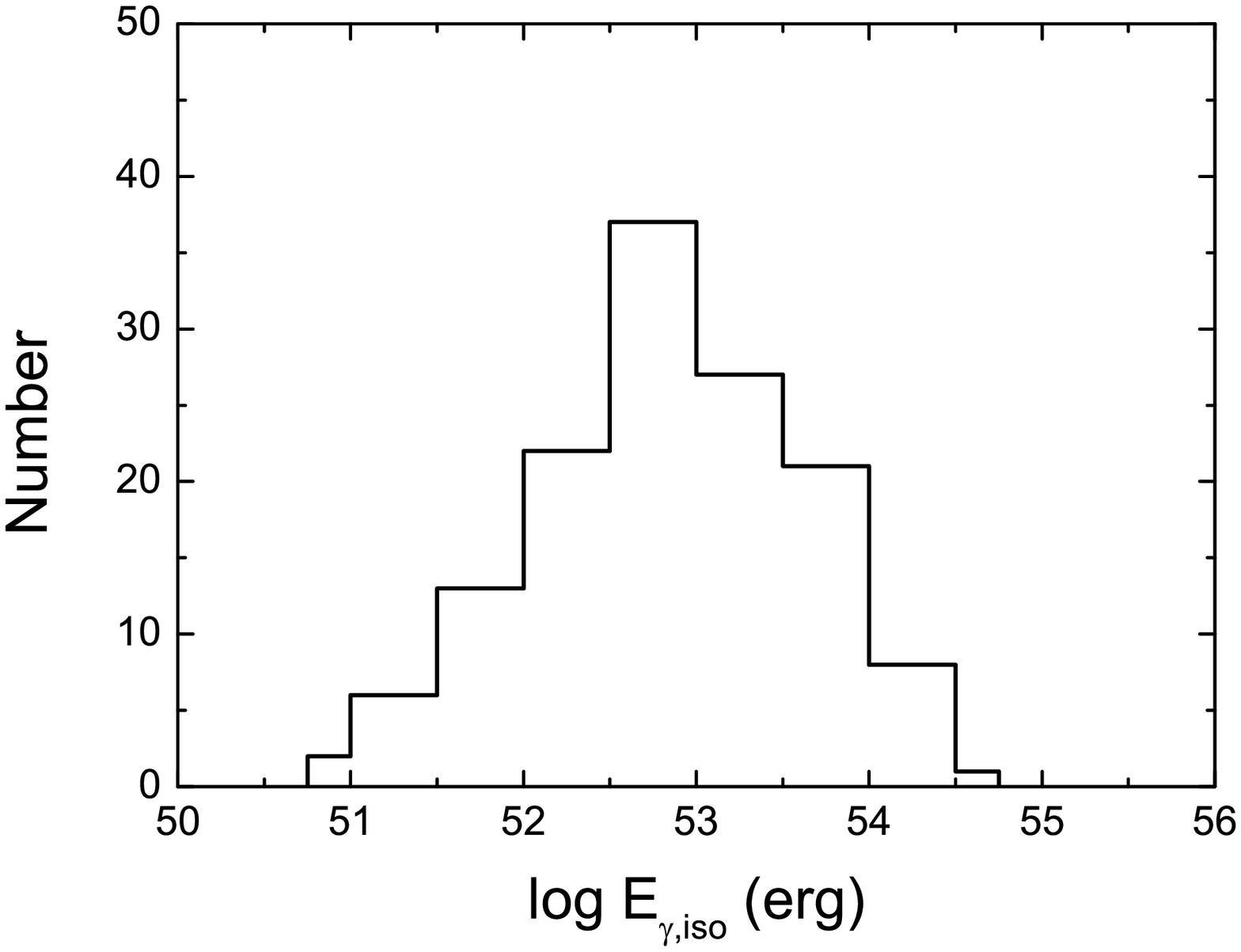}
\includegraphics[angle=0,scale=0.35]{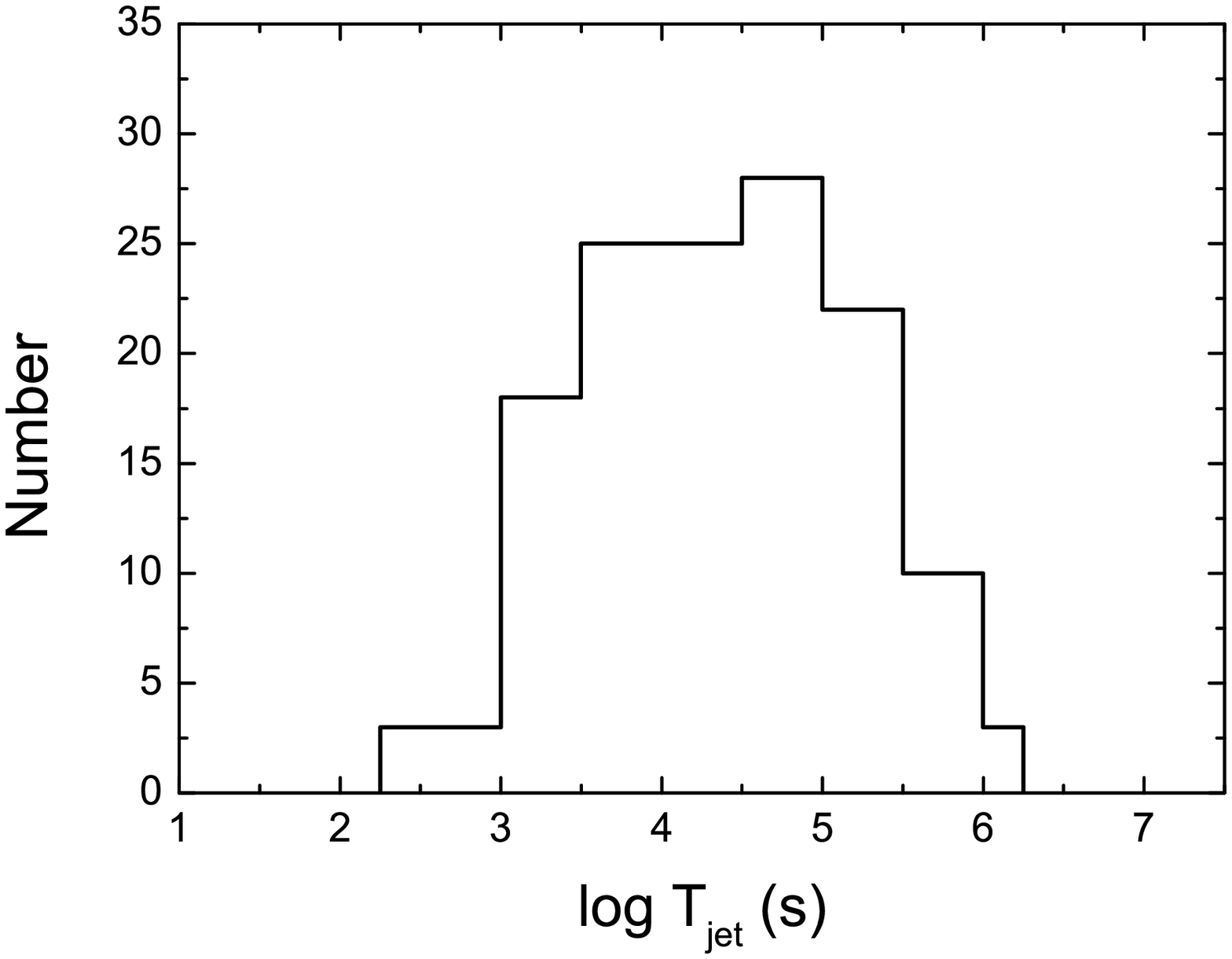}
\includegraphics[angle=0,scale=0.35]{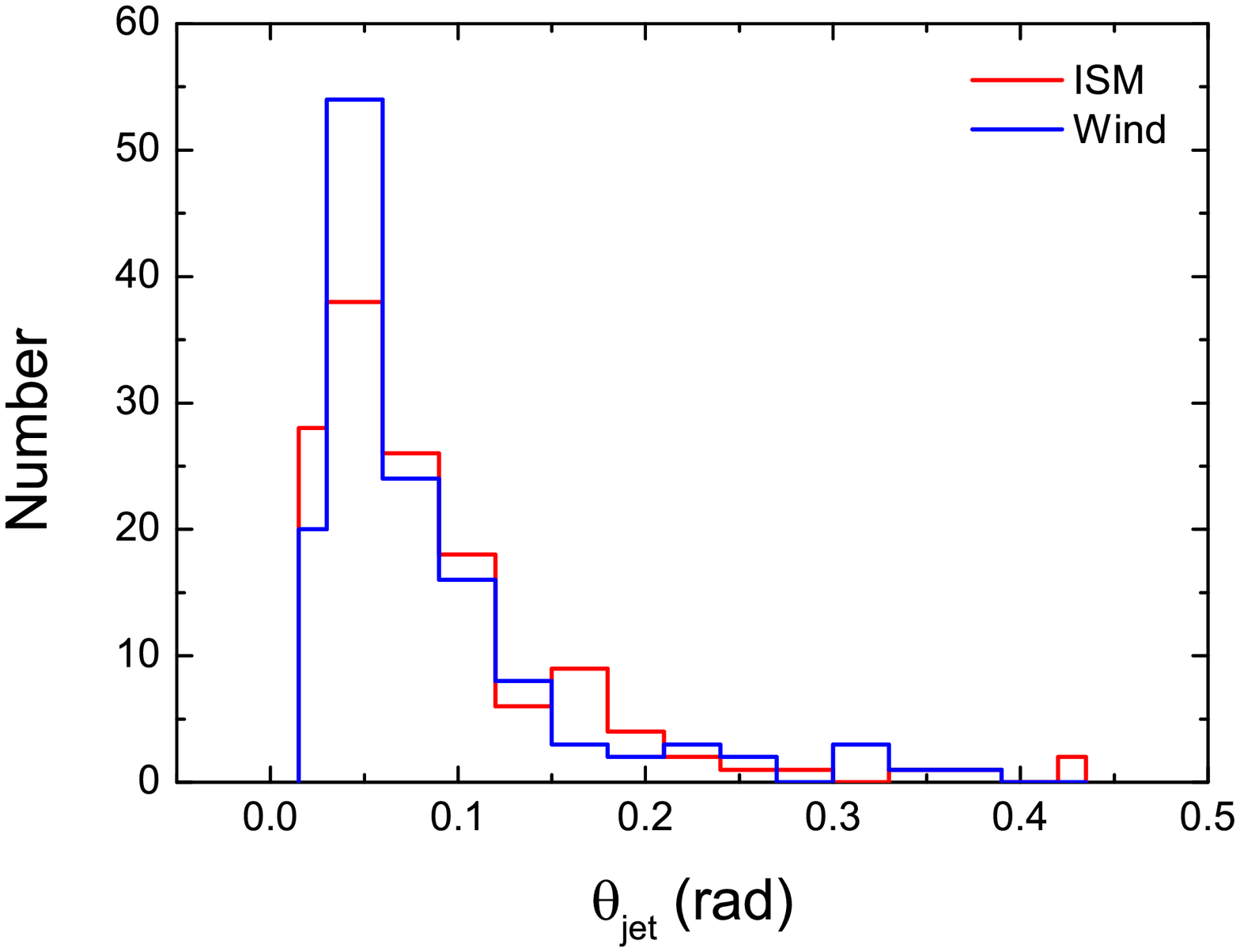}
\includegraphics[angle=0,scale=0.35]{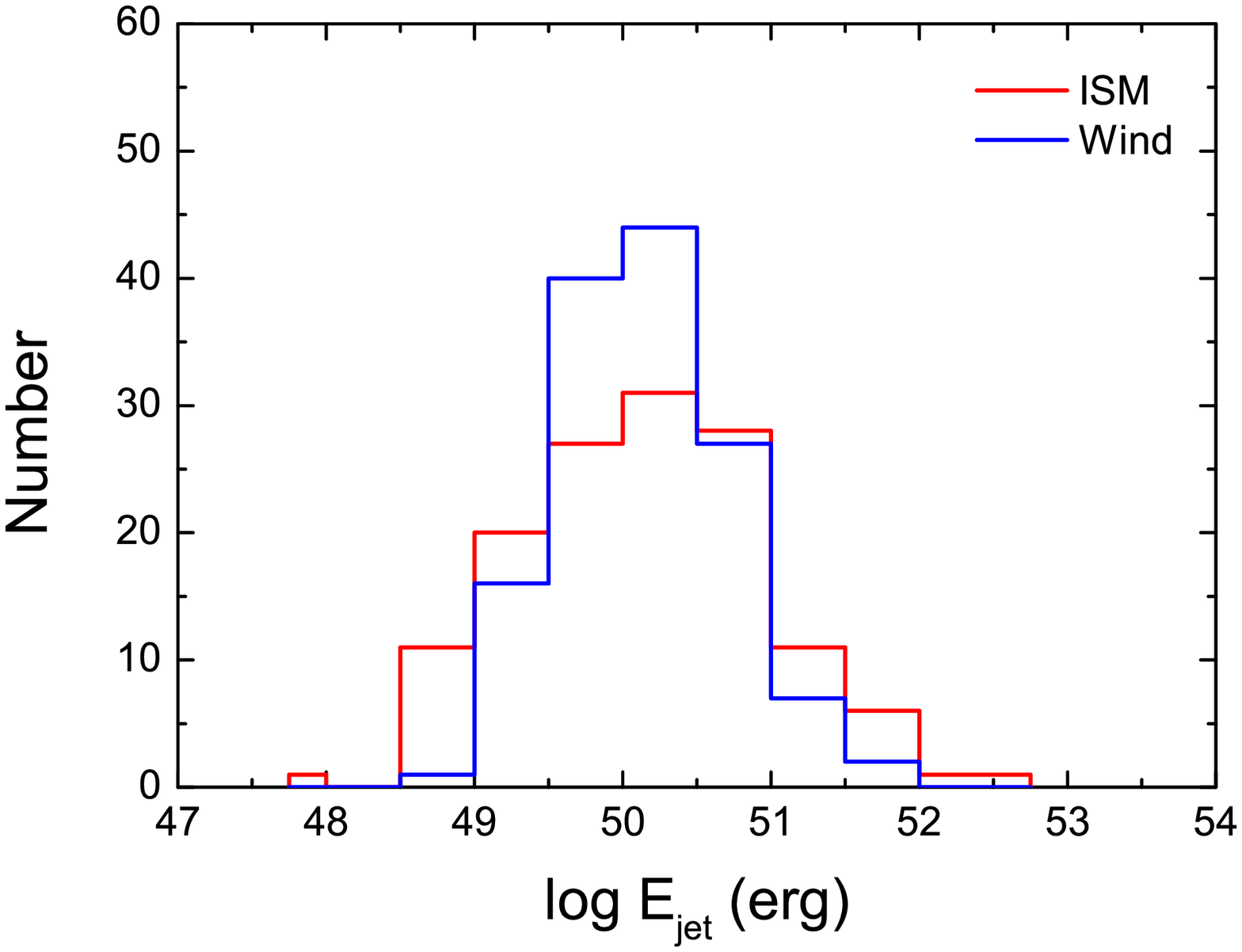}
\caption{The distributions of the parameters for GRBs with jet breaks.}
\end{figure*}

\begin{figure*}
\includegraphics[angle=0,scale=0.35]{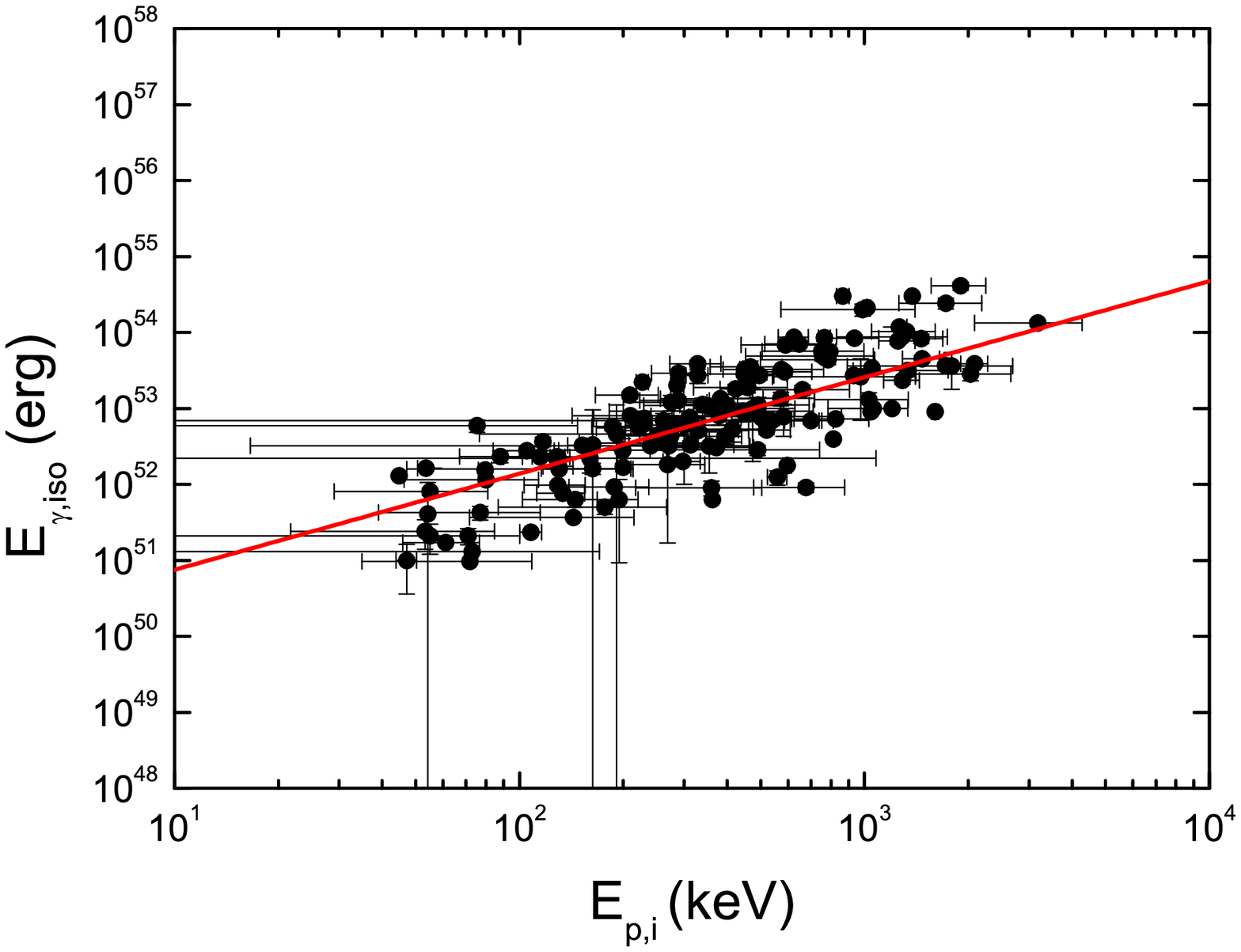}
\includegraphics[angle=0,scale=0.35]{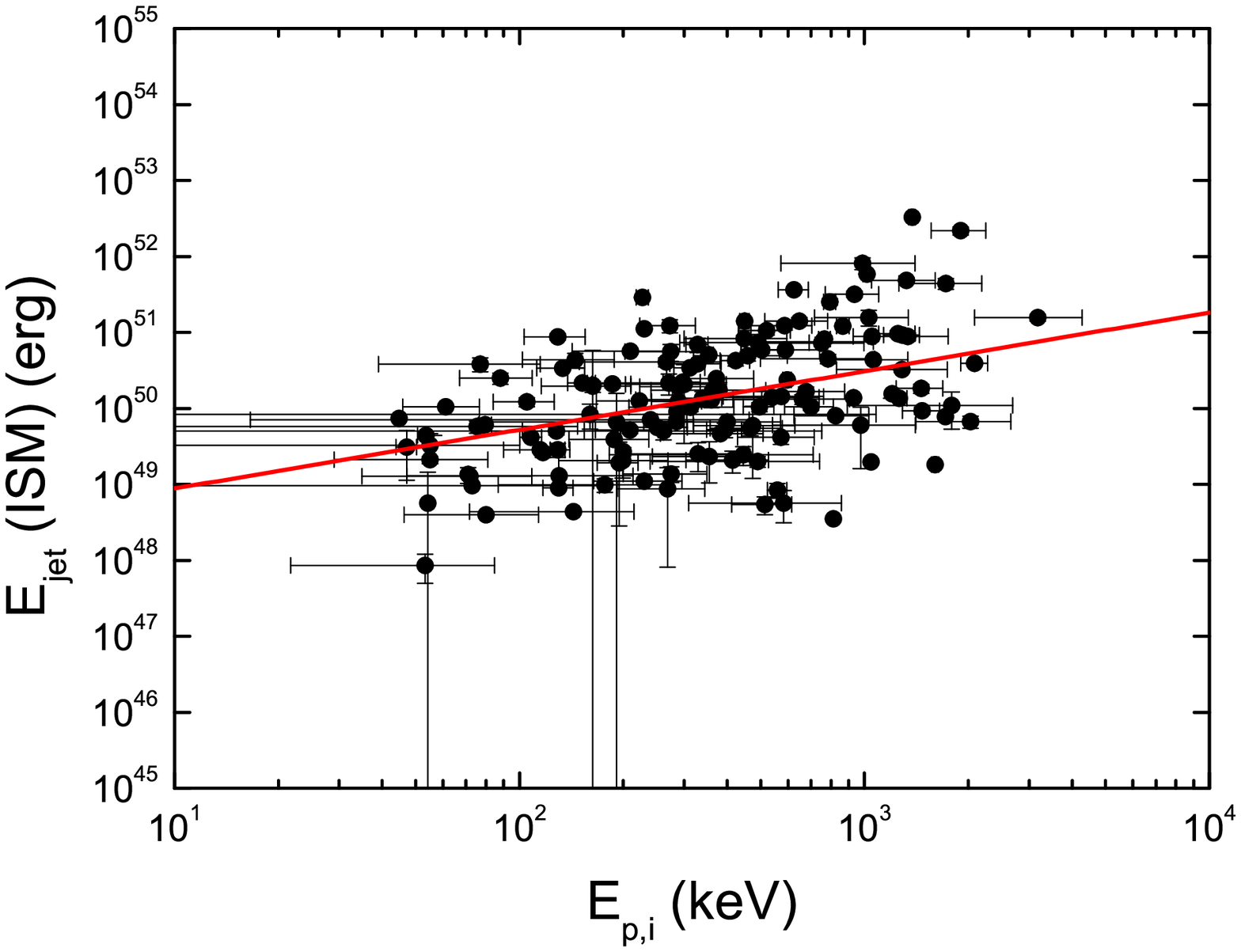}
\includegraphics[angle=0,scale=0.35]{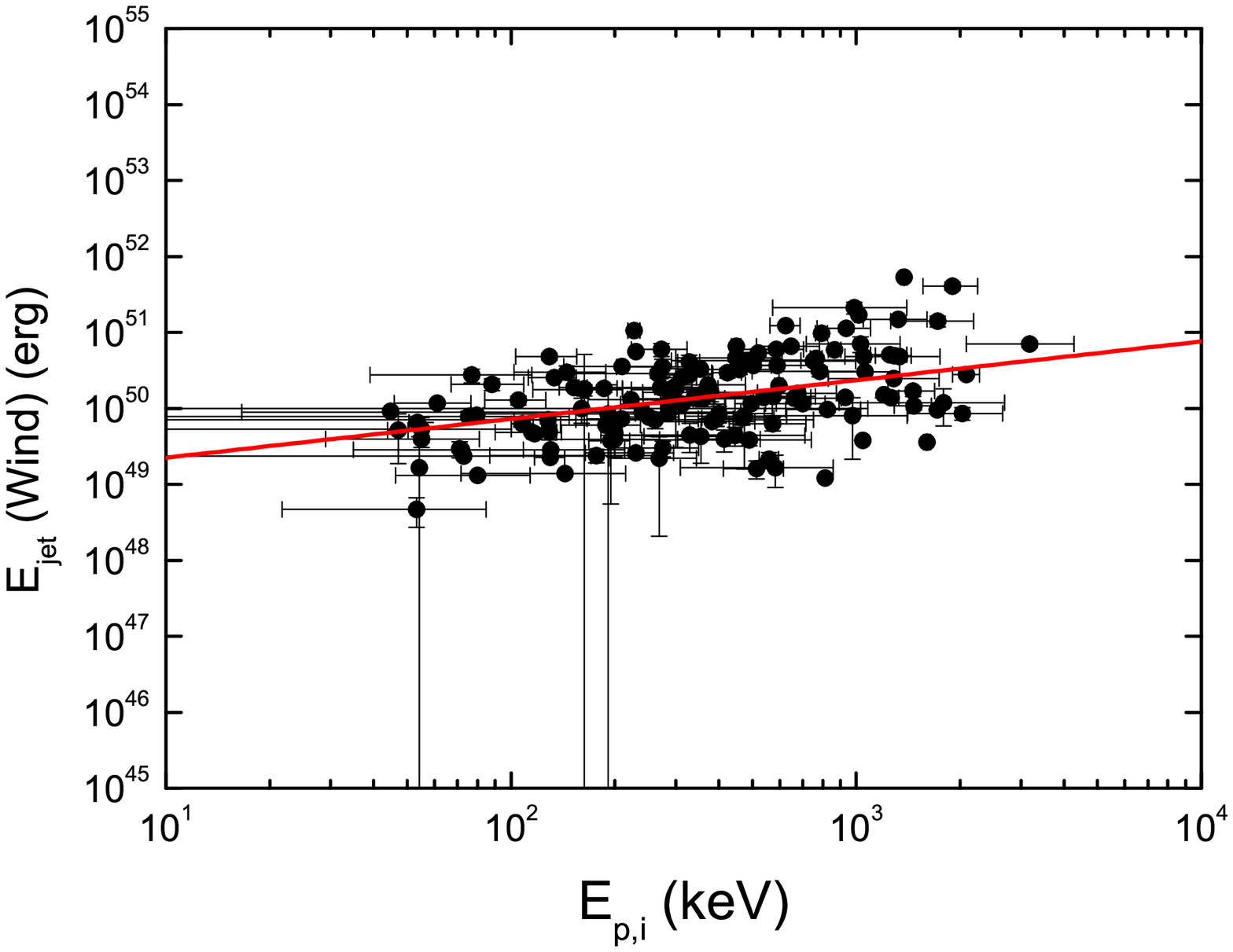}
\includegraphics[angle=0,scale=0.35]{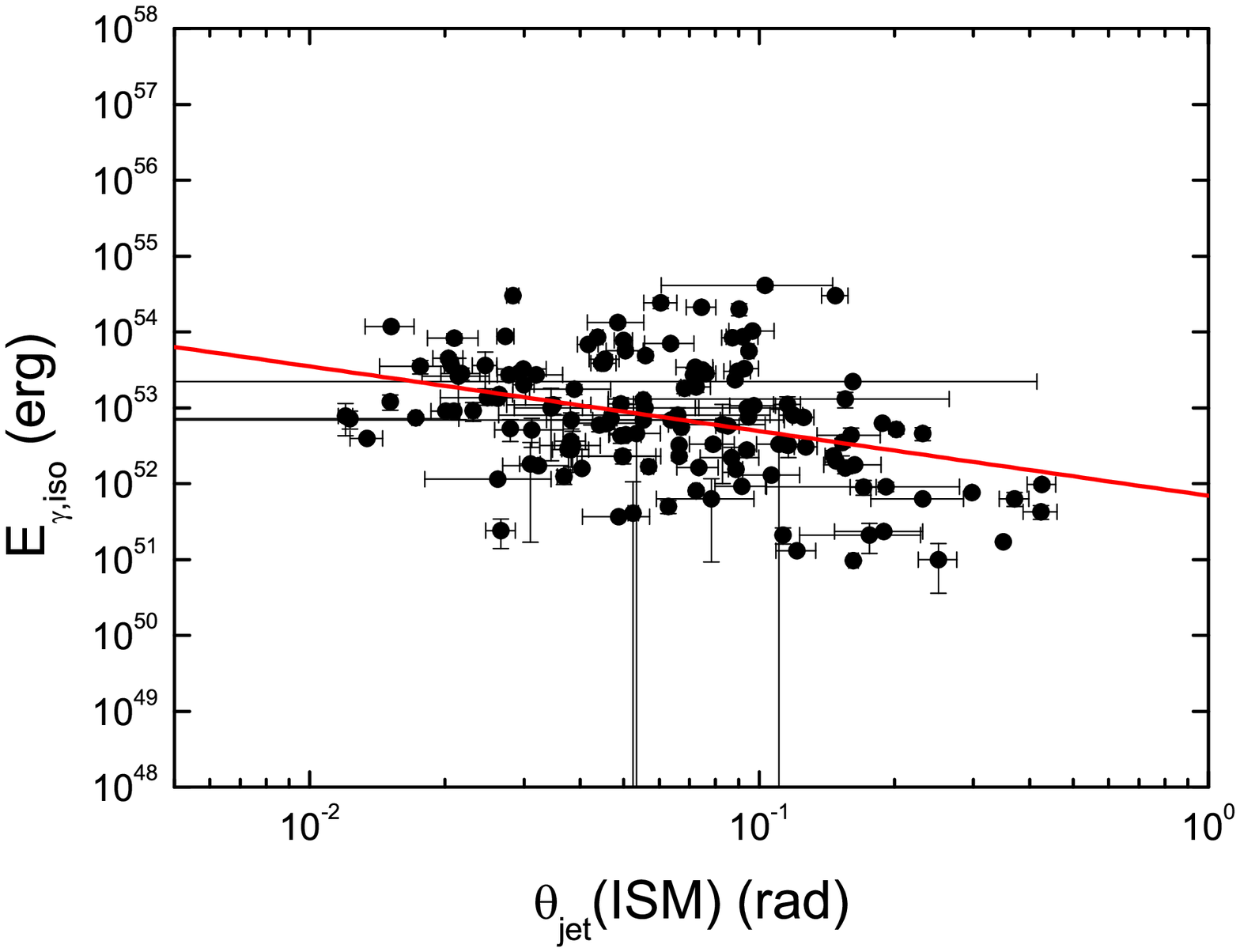}
\includegraphics[angle=0,scale=0.35]{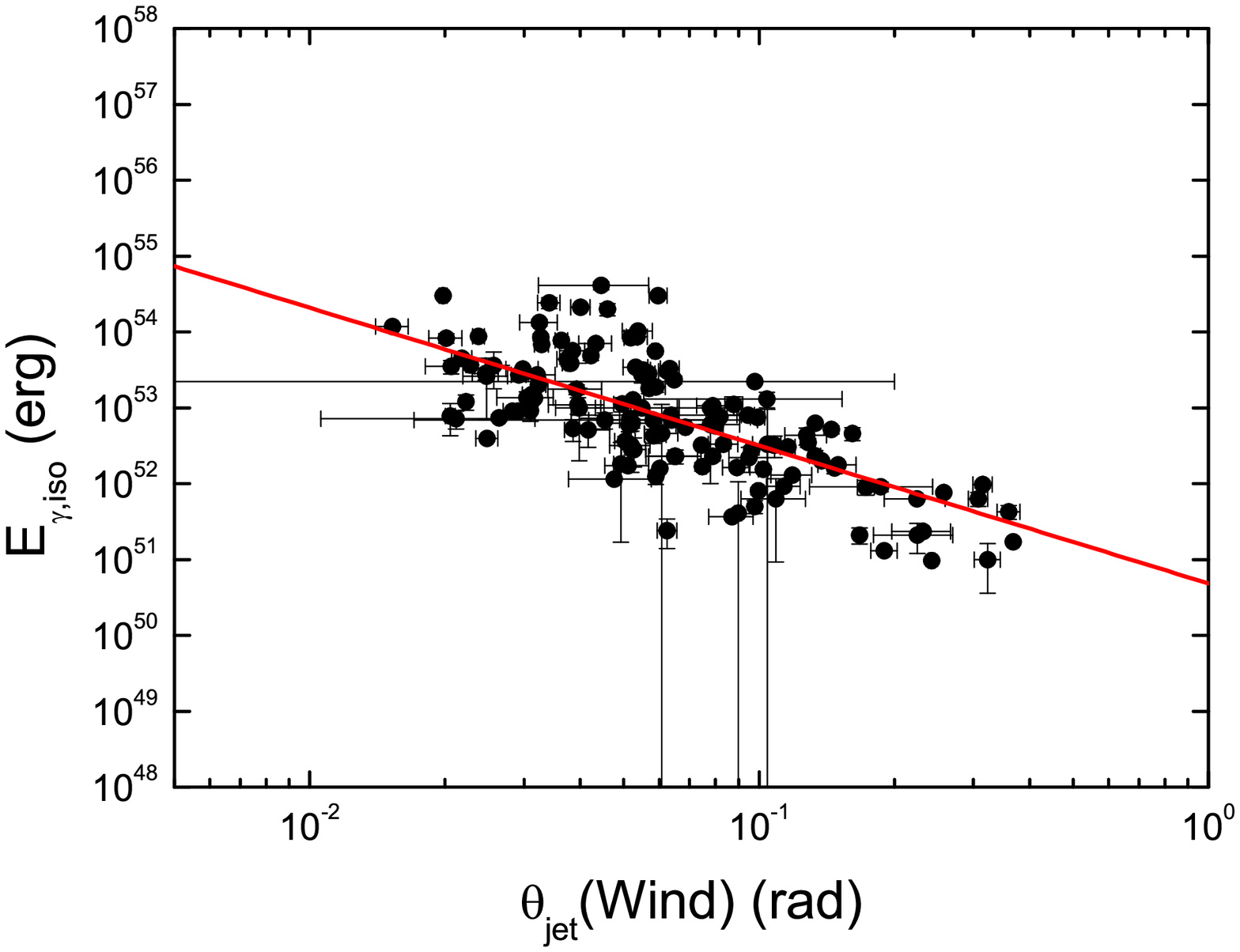}
\caption{The two-dimensional correlations of GRBs claimed in previous papers.}
\end{figure*}

\begin{figure*}
\includegraphics[angle=0,scale=0.33]{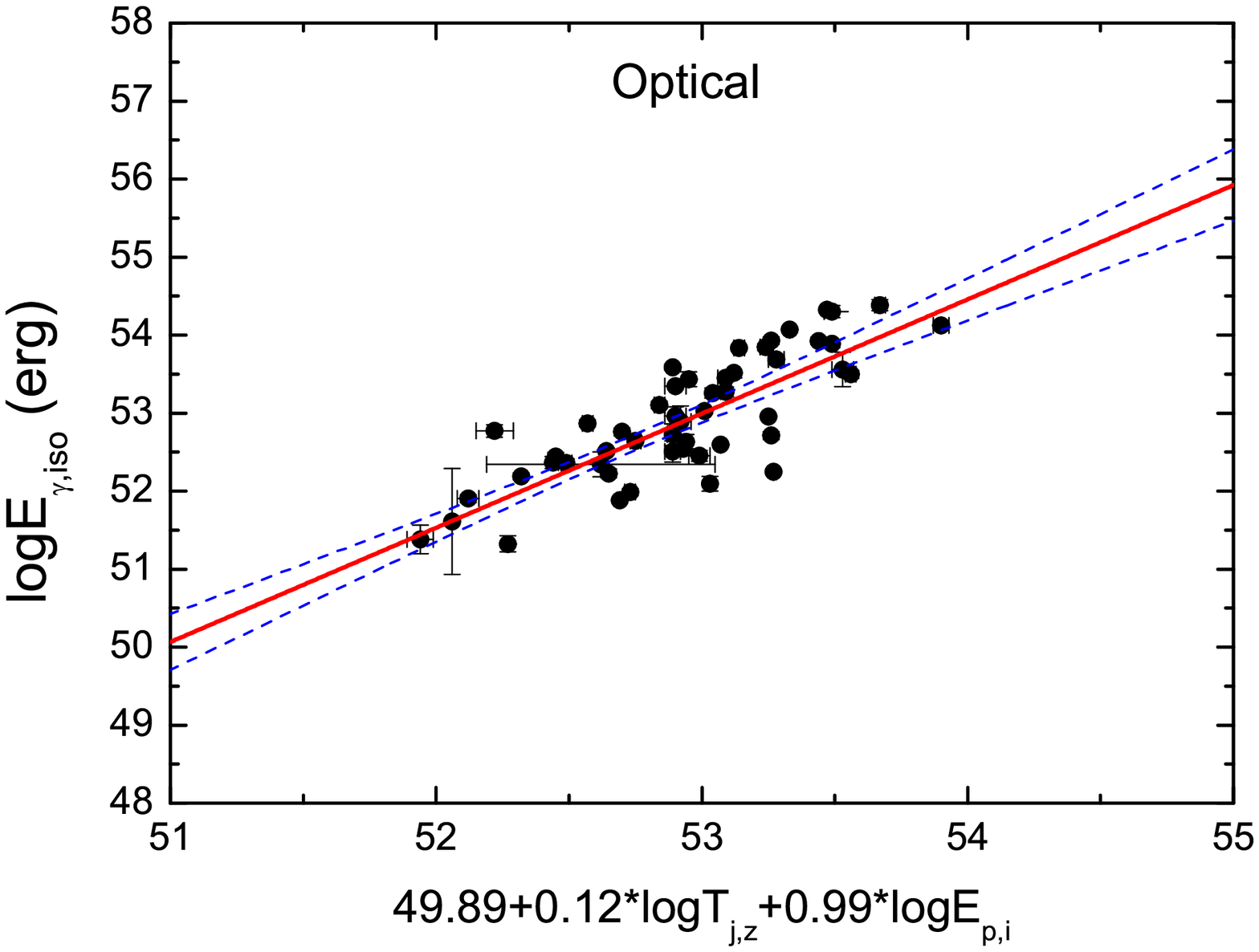}
\includegraphics[angle=0,scale=0.33]{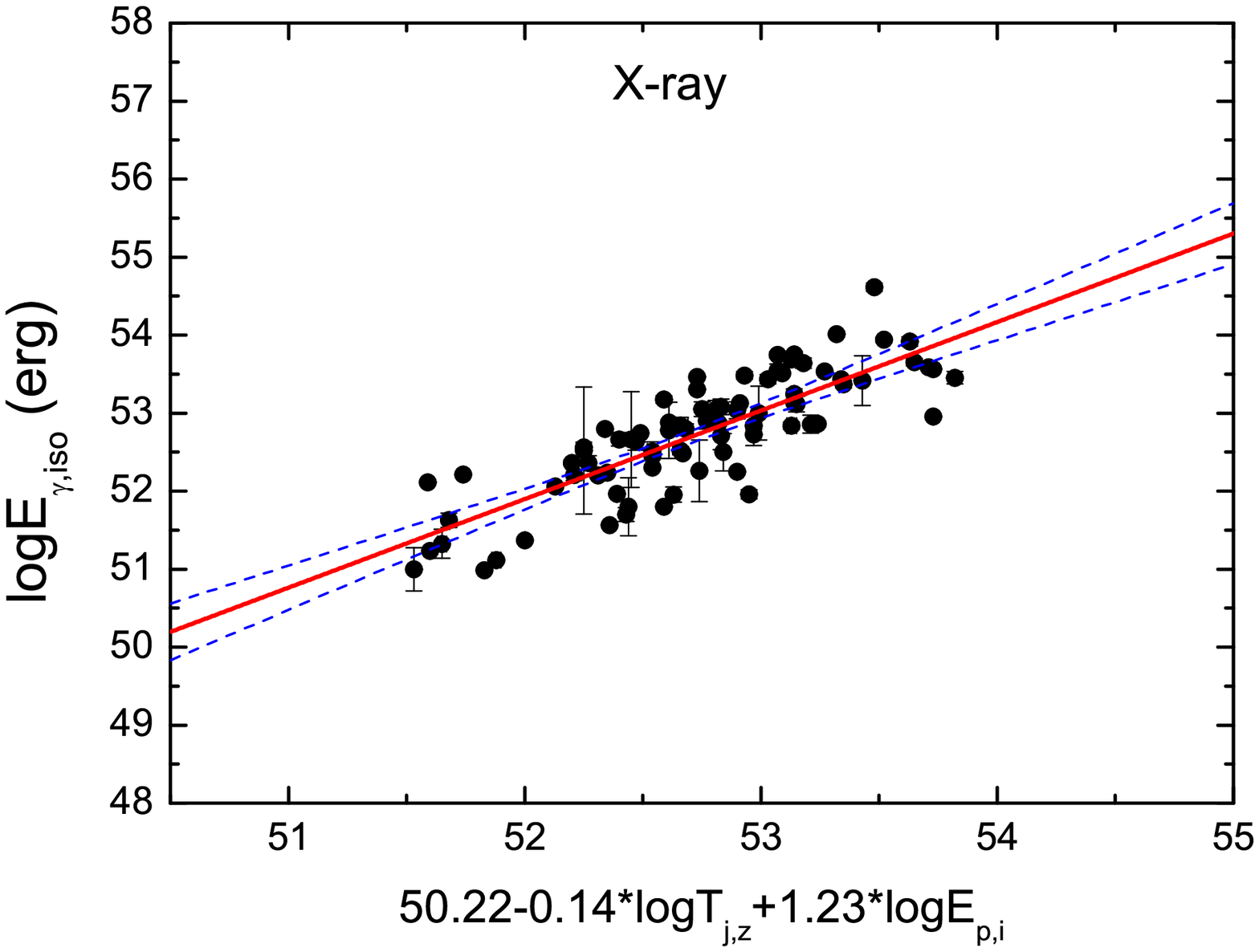}
\includegraphics[angle=0,scale=0.33]{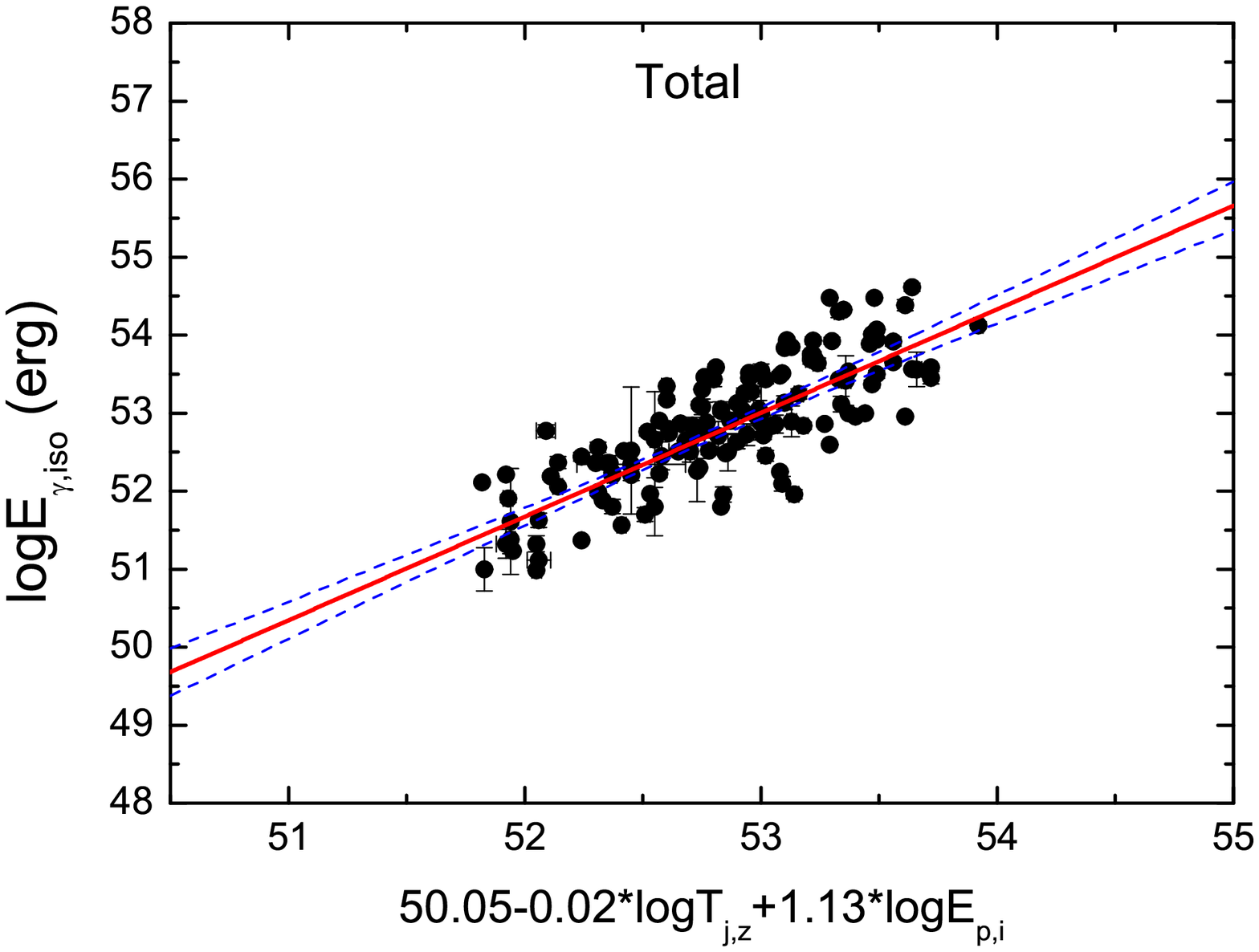}
\caption{The three-parameter correlation about $E_{\rm \gamma,iso}-T_{\rm j,z}-E_{\rm p,i}$ for different frequency data sample.}
\end{figure*}

\begin{figure*}
\includegraphics[angle=0,scale=0.33]{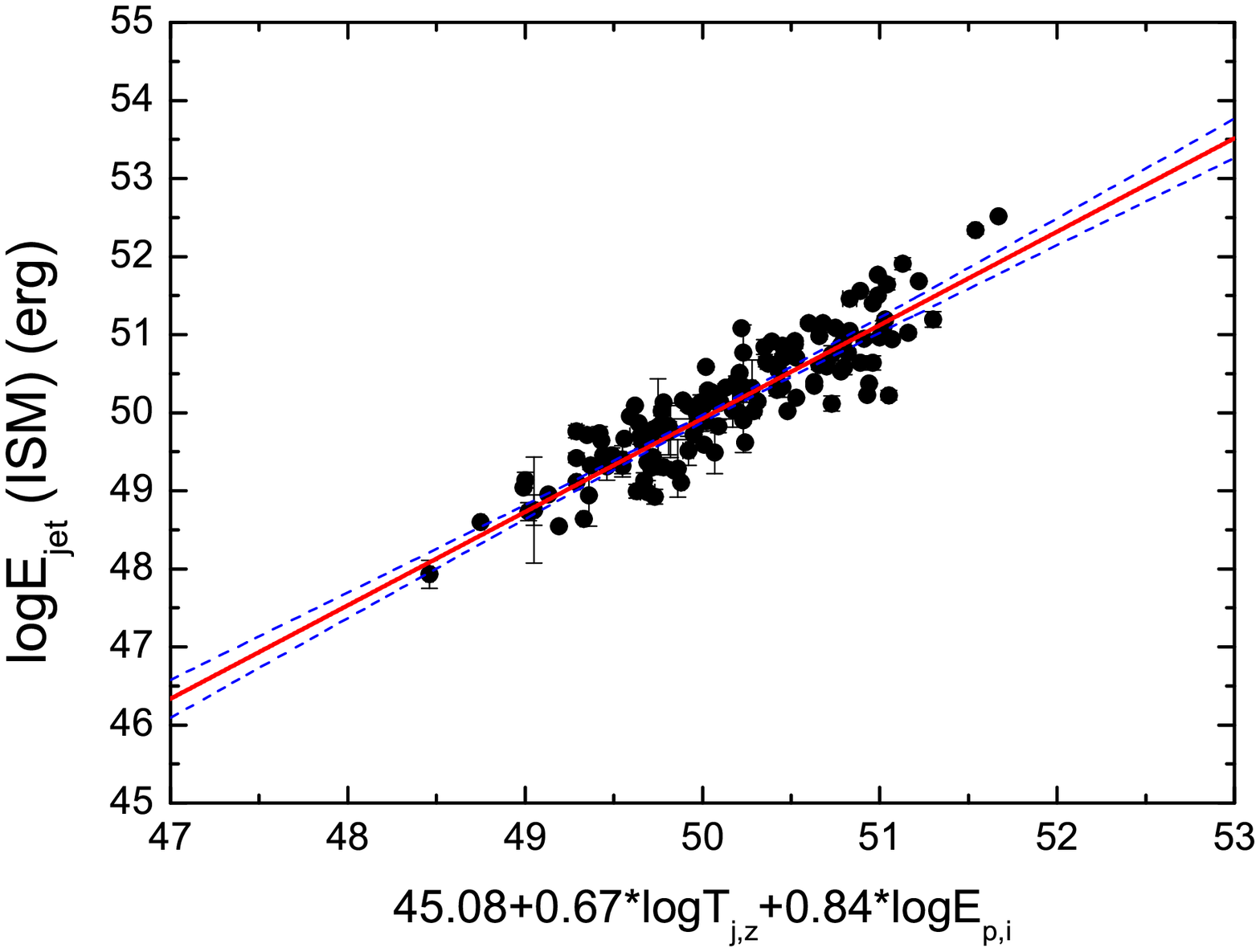}
\includegraphics[angle=0,scale=0.33]{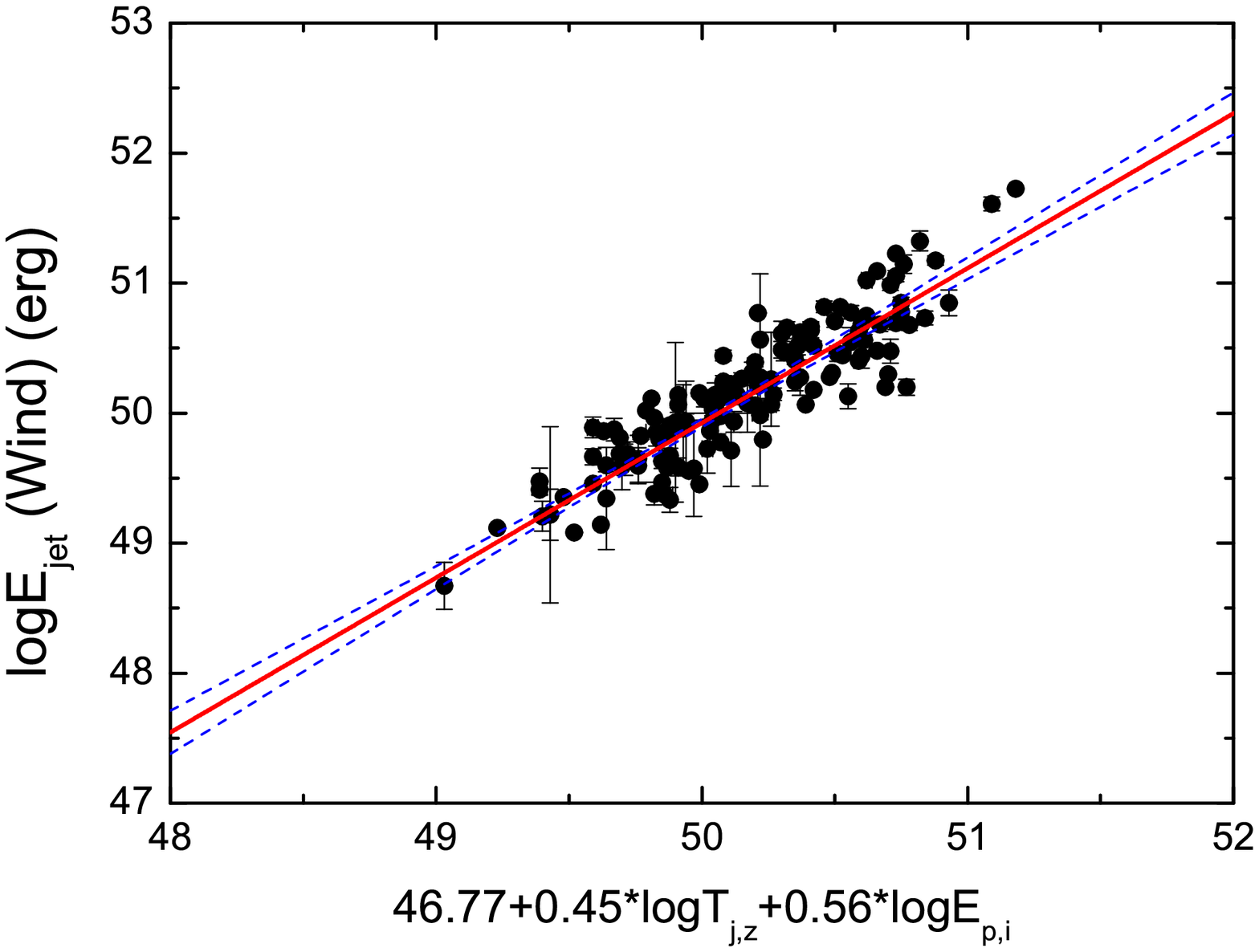}
\caption{Two new three-parameter correlations about $E_{\rm jet}-T_{\rm j,z}-E_{\rm p,i}$ for different circumburst mediums.}
\end{figure*}

%&&&&&&&&&&&&&&&&&&&&&&&&&&&&&&&&&&&&&&&&&&&&&&&&&&&&&&&&&&&&&&&&&&&&&&&&&&&&&&&&&&&&&&&&&&&&&&&&&&&&&&&&&&

\clearpage
\begin{deluxetable}{ccccccccccccccccccccccccc}
%\rotate
\tabletypesize{\scriptsize}
\tablecaption{The parameters of GRBs with jet breaks. }
\tablewidth{0pt}
\tabletypesize{\tiny}

\tablehead{ \colhead{GRB}
&\colhead{$z$}
&\colhead{$E_{\gamma,iso}$}
&\colhead{$E_{p,i}$}
&\colhead{$T_{jet} $}
&\colhead{$ Band $\tablenotemark{a}}
&\colhead{$\theta_{jet} (ISM)$}
&\colhead{$\theta_{jet} (Wind)$}
&\colhead{$E_{jet}(ISM)$}
&\colhead{$E_{jet}(Wind)$}
&\colhead{Refs.\tablenotemark{b}}\\
& &[10$^{52}$ erg] & [keV]&[day]&&[rad]&[rad]&[10$^{50}$ erg]&[10$^{50}$ erg]&}

\startdata
970508	&	0.8349	&	0.63	$\pm$	0.13	&	145	$\pm$	43	&	25	$\pm$	5	&	R	&	0.371	$\pm$	0.028	&	0.307	$\pm$	 0.015	&	4.34	$\pm$	0.92	&	2.99	$\pm$	0.63	&	1, 1, 2	\\
970828	&	0.9578	&	30.38	$\pm$	3.57	&	586	$\pm$	117	&	2.2	$\pm$	0.4	&	X	&	0.09	$\pm$	0.006	&	0.063	$\pm$	 0.003	&	12.2	$\pm$	1.43	&	5.95	$\pm$	0.7	&	1, 1, 2	\\
980703	&	0.9662	&	7.42	$\pm$	0.71	&	503	$\pm$	64	&	3.4	$\pm$	0.5	&	X	&	0.126	$\pm$	0.007	&	0.099	$\pm$	 0.004	&	5.85	$\pm$	0.56	&	3.65	$\pm$	0.35	&	1, 1, 2	\\
990123	&	1.6004	&	240.7	$\pm$	38.91	&	1724	$\pm$	466	&	2.04	$\pm$	0.46	&	O	&	0.06	$\pm$	0.005	&	0.034	 $\pm$	0.002	&	44	$\pm$	7.11	&	13.99	$\pm$	2.26	&	1, 1, 2	\\
990510	&	1.6187	&	18.1	$\pm$	2.72	&	423	$\pm$	42	&	1.2	$\pm$	0.08	&	O/R	&	0.068	$\pm$	0.002	&	0.057	$\pm$	 0.001	&	4.22	$\pm$	0.63	&	2.93	$\pm$	0.44	&	1, 1, 2	\\
990705	&	0.8424	&	18.7	$\pm$	2.67	&	459	$\pm$	139	&	1	$\pm$	0.2	&	O	&	0.072	$\pm$	0.005	&	0.059	$\pm$	 0.003	&	4.91	$\pm$	0.7	&	3.24	$\pm$	0.46	&	1, 1, 2	\\
990712	&	0.434	&	0.76	$\pm$	0.04	&	133.4	$\pm$	21.3	&	11.57			&	O	&	0.298			&	0.258			&	 3.37	$\pm$	0.18	&	2.52	$\pm$	0.13	&	3, 3, 3	\\
991216	&	1.02	&	69.79	$\pm$	7.16	&	648	$\pm$	134	&	1.2	$\pm$	0.4	&	O	&	0.064	$\pm$	0.008	&	0.043	$\pm$	 0.004	&	14.11	$\pm$	1.45	&	6.56	$\pm$	0.67	&	1, 1, 2	\\
000301C	&	2.03	&	199	$\pm$	35	&	987.8	$\pm$	415.1	&	6.52	$\pm$	0.22	&	O	&	0.09	$\pm$	0.001	&	0.046			 &	81.26	$\pm$	14.29	&	21.06	$\pm$	3.7	&	3, 3, 3	\\
000926A	&	2.0379	&	27.1	$\pm$	5.9	&	328	$\pm$	24	&	1.8	$\pm$	0.1	&	O	&	0.072	$\pm$	0.001	&	0.055	$\pm$	0.001	 &	6.93	$\pm$	1.51	&	4.08	$\pm$	0.89	&	4, 5, 6	\\
010222A	&	1.477	&	84.9	$\pm$	9.03	&	766	$\pm$	30	&	0.58	$\pm$	0.04	&	O	&	0.044	$\pm$	0.001	&	0.033	 $\pm$	0.001	&	8.13	$\pm$	0.87	&	4.54	$\pm$	0.48	&	1, 1, 7	\\
010921	&	0.4509	&	0.97	$\pm$	0.1	&	129	$\pm$	26	&	33	$\pm$	6.5	&	O	&	0.426	$\pm$	0.031	&	0.314	$\pm$	0.015	 &	8.77	$\pm$	0.88	&	4.78	$\pm$	0.48	&	1, 1, 2	\\
011121A	&	0.36	&	9.89	$\pm$	0.27	&	1060	$\pm$	275	&	1.2	$\pm$	0.75	&	R	&	0.094	$\pm$	0.022	&	0.078	 $\pm$	0.012	&	4.39	$\pm$	0.12	&	3.01	$\pm$	0.08	&	5, 1, 8	\\
011211	&	2.14	&	5.74	$\pm$	0.64	&	186	$\pm$	24	&	1.77	$\pm$	0.28	&	O	&	0.085	$\pm$	0.005	&	0.08	 $\pm$	0.003	&	2.08	$\pm$	0.23	&	1.83	$\pm$	0.2	&	1, 1, 2	\\
020124	&	3.198	&	28.46	$\pm$	2.75	&	448	$\pm$	148	&	3	$\pm$	0.4	&	O	&	0.076	$\pm$	0.004	&	0.057	$\pm$	 0.002	&	8.27	$\pm$	0.8	&	4.59	$\pm$	0.44	&	1, 1, 2	\\
020405	&	0.6899	&	10.64	$\pm$	0.89	&	354	$\pm$	10	&	1.67	$\pm$	0.52	&	O	&	0.097	$\pm$	0.011	&	0.079	 $\pm$	0.006	&	5.04	$\pm$	0.42	&	3.3	$\pm$	0.28	&	1, 1, 2	\\
020813	&	1.254	&	68.35	$\pm$	1.71	&	590	$\pm$	151	&	0.43	$\pm$	0.06	&	O	&	0.042	$\pm$	0.002	&	0.033	 $\pm$	0.001	&	5.93	$\pm$	0.15	&	3.68	$\pm$	0.09	&	1, 1, 2	\\
021004	&	2.332	&	3.47	$\pm$	0.46	&	266	$\pm$	117	&	7.6	$\pm$	0.3	&	O	&	0.153	$\pm$	0.002	&	0.129	$\pm$	 0.001	&	4.07	$\pm$	0.54	&	2.86	$\pm$	0.38	&	1, 1, 2	\\
030226A	&	1.986	&	12.73	$\pm$	1.36	&	289	$\pm$	66	&	0.69	$\pm$	0.04	&	O	&	0.055	$\pm$	0.001	&	0.052	 $\pm$	0.001	&	1.94	$\pm$	0.21	&	1.75	$\pm$	0.19	&	1, 1, 9	\\
030323	&	3.37	&	3.2	$\pm$	1	&	270.9	$\pm$	113.6	&	4.63			&	O	&	0.116			&	0.108			&	2.16	 $\pm$	0.67	&	1.87	$\pm$	0.59	&	3, 3, 3	\\
030328A	&	1.5216	&	38.86	$\pm$	3.62	&	328	$\pm$	55	&	0.48	$\pm$	0.03	&	O	&	0.045	$\pm$	0.001	&	0.038	 $\pm$	0.001	&	3.87	$\pm$	0.36	&	2.77	$\pm$	0.26	&	1, 1, 10	\\
030329	&	0.1685	&	1.55	$\pm$	0.15	&	79.5	$\pm$	3.5	&	0.47	$\pm$	0.05	&	O	&	0.089	$\pm$	0.003	&	0.102	 $\pm$	0.002	&	0.61	$\pm$	0.06	&	0.81	$\pm$	0.08	&	3, 3, 3	\\
030429A	&	2.658	&	2.29	$\pm$	0.27	&	128	$\pm$	26	&	0.78	$\pm$	0.01	&	O	&	0.066			&	0.079			&	 0.5	$\pm$	0.06	&	0.71	$\pm$	0.09	&	1, 1, 11	\\
050315	&	1.949	&	3.3	$\pm$	6.2	&	163			&	2.78	$\pm$	0.8	&	X	&	0.111	$\pm$	0.012	&	0.104	$\pm$	0.008	&	 2.02	$\pm$	3.8	&	1.8	$\pm$	3.37	&	12, 12, 2	\\
050318A	&	1.44	&	2.3	$\pm$	0.16	&	115	$\pm$	25	&	0.24	$\pm$	0.12	&	X	&	0.05	$\pm$	0.009	&	0.065	$\pm$	 0.008	&	0.28	$\pm$	0.02	&	0.48	$\pm$	0.03	&	1, 1, 2	\\
050319	&	3.2425	&	4.6	$\pm$	6.5	&	190.9	$\pm$	114.5	&	0.64	$\pm$	0.22	&	X	&	0.053	$\pm$	0.007	&	0.061	 $\pm$	0.005	&	0.66	$\pm$	0.93	&	0.85	$\pm$	1.2	&	13, 13, 2	\\
050401	&	2.9	&	35	$\pm$	7	&	467	$\pm$	110	&	0.06	$\pm$	0.03	&	X	&	0.018	$\pm$	0.003	&	0.021	$\pm$	0.003	 &	0.54	$\pm$	0.11	&	0.75	$\pm$	0.15	&	13, 13, 14	\\
050408	&	1.236	&	1.3			&	44.7			&	1.39	$\pm$	0.58	&	X	&	0.106	$\pm$	0.017	&	0.119	$\pm$	0.012	 &	0.74			&	0.92			&	15, 15, 2	\\
050502A	&	3.793	&	9.12	$\pm$	2.52	&	445.7	$\pm$	263.6	&	0.1	$\pm$	0.01	&	O	&	0.023	$\pm$	0.001	&	0.031			 &	0.24	$\pm$	0.07	&	0.44	$\pm$	0.12	&	16, 16, 3	\\
050505A	&	4.2748	&	17.6	$\pm$	2.61	&	661	$\pm$	245	&	0.53	$\pm$	0.29	&	X	&	0.039	$\pm$	0.008	&	0.039	 $\pm$	0.005	&	1.32	$\pm$	0.2	&	1.35	$\pm$	0.2	&	5, 5, 2	\\
050525A	&	0.606	&	2.3	$\pm$	0.49	&	129	$\pm$	6.5	&	0.16	$\pm$	0.09	&	X	&	0.05	$\pm$	0.011	&	0.065	$\pm$	 0.009	&	0.29	$\pm$	0.06	&	0.49	$\pm$	0.1	&	1, 1, 2	\\
050730A	&	3.969	&	26	$\pm$	19	&	974.1	$\pm$	432.4	&	0.12	$\pm$	0.05	&	X	&	0.021	$\pm$	0.004	&	0.025	 $\pm$	0.003	&	0.6	$\pm$	0.44	&	0.8	$\pm$	0.58	&	17, 13, 17	\\
050801	&	1.56	&	0.41	$\pm$	0.64	&	54.3			&	0.16	$\pm$	0.01	&	O	&	0.052	$\pm$	0.001	&	0.09	 $\pm$	0.002	&	0.06	$\pm$	0.09	&	0.17	$\pm$	0.26	&	18, 19, 3	\\
050802	&	1.71	&	1.82	$\pm$	1.65	&	268.3	$\pm$	75.9	&	0.07	$\pm$	0.01	&	X	&	0.031	$\pm$	0.002	&	 0.049	$\pm$	0.002	&	0.09	$\pm$	0.08	&	0.22	$\pm$	0.2	&	13, 13, 2	\\
050814	&	5.3	&	11.2	$\pm$	2.43	&	339	$\pm$	47	&	1.03	$\pm$	0.18	&	X	&	0.049	$\pm$	0.003	&	0.05	$\pm$	 0.002	&	1.36	$\pm$	0.29	&	1.38	$\pm$	0.3	&	5, 5, 2	\\
050820A	&	2.62	&	103.36	$\pm$	8.23	&	1325	$\pm$	277	&	7.52	$\pm$	2.31	&	X	&	0.097	$\pm$	0.011	&	0.054	 $\pm$	0.004	&	48.45	$\pm$	3.86	&	14.92	$\pm$	1.19	&	1, 1, 2	\\
050904	&	6.29	&	133.36	$\pm$	13.89	&	3178	$\pm$	1094	&	2.6	$\pm$	1	&	O	&	0.048	$\pm$	0.007	&	0.032	 $\pm$	0.003	&	15.64	$\pm$	1.63	&	7.02	$\pm$	0.73	&	1, 5, 2	\\
050922C	&	2.198	&	5.3	$\pm$	1.7	&	415	$\pm$	111	&	0.09			&	O/X	&	0.028			&	0.039			&	0.21	$\pm$	 0.07	&	0.39	$\pm$	0.13	&	5,  5, 2	\\
051016B	&	0.94	&	0.1			&	71.8	$\pm$	36.9	&	1.56	$\pm$	0.1	&	X	&	0.162	$\pm$	0.004	&	0.242	$\pm$	 0.004	&	0.13			&	0.28			&	2, 20, 2	\\
051022	&	0.8	&	56.04	$\pm$	5.33	&	794	$\pm$	32	&	2.9	$\pm$	0.2	&	X	&	0.095	$\pm$	0.002	&	0.059	$\pm$	0.001	 &	25.3	$\pm$	2.41	&	9.68	$\pm$	0.92	&	1, 5, 2	\\
051109A	&	2.35	&	6.85	$\pm$	0.73	&	539	$\pm$	200	&	0.92	$\pm$	0.71	&	X	&	0.064	$\pm$	0.018	&	0.064	 $\pm$	0.012	&	1.39	$\pm$	0.15	&	1.4	$\pm$	0.15	&	1, 1, 2	\\
051221A	&	0.5465	&	0.91	$\pm$	0.13	&	677	$\pm$	200	&	4.1	$\pm$	5	&	X	&	0.192	$\pm$	0.088	&	0.186	$\pm$	 0.057	&	1.67	$\pm$	0.24	&	1.58	$\pm$	0.22	&	5, 5, 21	\\
060115	&	3.53	&	6.3	$\pm$	0.9	&	285	$\pm$	34	&	0.51	$\pm$	0.22	&	X	&	0.046	$\pm$	0.007	&	0.052	$\pm$	 0.006	&	0.67	$\pm$	0.1	&	0.86	$\pm$	0.12	&	5, 5, 2	\\
060124	&	2.3	&	43.79	$\pm$	6.39	&	784	$\pm$	285	&	0.68	$\pm$	0.14	&	X	&	0.045	$\pm$	0.003	&	0.037	$\pm$	 0.002	&	4.5	$\pm$	0.66	&	3.06	$\pm$	0.45	&	1, 1, 2	\\
060206A	&	4.048	&	4.3	$\pm$	0.9	&	394	$\pm$	46	&	0.6			&	O	&	0.049			&	0.058			&	0.52	$\pm$	 0.11	&	0.73	$\pm$	0.15	&	5, 5, 22	\\
060210	&	3.91	&	32.23	$\pm$	1.84	&	575	$\pm$	186	&	0.3	$\pm$	0.1	&	X	&	0.03	$\pm$	0.004	&	0.03	$\pm$	 0.002	&	1.44	$\pm$	0.08	&	1.43	$\pm$	0.08	&	1, 5, 2	\\
060418	&	1.49	&	13.55	$\pm$	2.71	&	572	$\pm$	143	&	0.07	$\pm$	0.04	&	X	&	0.025	$\pm$	0.005	&	0.03	 $\pm$	0.004	&	0.42	$\pm$	0.08	&	0.63	$\pm$	0.13	&	1, 5, 2	\\
060526	&	3.22	&	2.75	$\pm$	0.37	&	105	$\pm$	21	&	2.41	$\pm$	0.06	&	O	&	0.094	$\pm$	0.001	&	0.096	 $\pm$	0.001	&	1.21	$\pm$	0.16	&	1.28	$\pm$	0.17	&	1, 1, 2	\\
060605A	&	3.773	&	2.83	$\pm$	0.45	&	490	$\pm$	251	&	0.24	$\pm$	0.02	&	O	&	0.038	$\pm$	0.001	&	0.052	 $\pm$	0.001	&	0.2	$\pm$	0.03	&	0.38	$\pm$	0.06	&	5, 5, 2	\\
060614	&	0.12	&	0.21	$\pm$	0.09	&	55	$\pm$	45	&	1.45	$\pm$	1.16	&	X	&	0.176	$\pm$	0.053	&	0.225	 $\pm$	0.045	&	0.33	$\pm$	0.14	&	0.53	$\pm$	0.23	&	13, 13, 2	\\
060707	&	3.425	&	4.32	$\pm$	1.1	&	274	$\pm$	72	&	12.26	$\pm$	5.25	&	X	&	0.16	$\pm$	0.026	&	0.128	$\pm$	 0.014	&	5.56	$\pm$	1.42	&	3.52	$\pm$	0.9	&	1, 1, 2	\\
060714	&	2.711	&	13.4	$\pm$	0.9	&	382.6	$\pm$	126.2	&	0.12	$\pm$	0.01	&	X	&	0.026	$\pm$	0.001	&	0.032	 $\pm$	0.001	&	0.46	$\pm$	0.03	&	0.67	$\pm$	0.05	&	5, 23, 2	\\
060729	&	0.54	&	0.42	$\pm$	0.09	&	77	$\pm$	38	&	26.23	$\pm$	6.12	&	X	&	0.424	$\pm$	0.037	&	0.359	 $\pm$	0.021	&	3.8	$\pm$	0.79	&	2.73	$\pm$	0.57	&	1, 1, 2	\\
060814	&	0.84	&	56.71	$\pm$	5.27	&	751	$\pm$	246	&	0.55	$\pm$	0.14	&	X	&	0.05	$\pm$	0.005	&	0.038	 $\pm$	0.002	&	7.22	$\pm$	0.67	&	4.19	$\pm$	0.39	&	1, 1, 2	\\
060906	&	3.69	&	14.9	$\pm$	1.56	&	209	$\pm$	43	&	0.16	$\pm$	0.03	&	X	&	0.026	$\pm$	0.002	&	0.031	 $\pm$	0.001	&	0.52	$\pm$	0.05	&	0.73	$\pm$	0.08	&	5, 5, 2	\\
060908	&	1.8836	&	7.18	$\pm$	1.91	&	514	$\pm$	102	&	0.01	$\pm$	0.02	&	X	&	0.012	$\pm$	0.009	&	0.021	 $\pm$	0.011	&	0.05	$\pm$	0.01	&	0.16	$\pm$	0.04	&	1, 13, 2	\\
060926	&	3.21	&	1.15			&	80	$\pm$	33.7	&	0.06	$\pm$	0.05	&	X	&	0.026	$\pm$	0.008	&	0.048	$\pm$	 0.01	&	0.04			&	0.13			&	2, 20, 2	\\
060927A	&	5.47	&	12.02	$\pm$	2.77	&	275	$\pm$	75	&	0.05			&	X	&	0.015			&	0.022			&	0.14	 $\pm$	0.03	&	0.3	$\pm$	0.07	&	1, 1, 24	\\
061121	&	1.314	&	23.5	$\pm$	2.7	&	1289	$\pm$	153	&	2.31			&	X	&	0.089			&	0.065			&	9.21	 $\pm$	1.06	&	4.93	$\pm$	0.57	&	1, 1, 2	\\
061126	&	1.1588	&	31.42	$\pm$	3.59	&	1337	$\pm$	410	&	1.52	$\pm$	0.23	&	O	&	0.075	$\pm$	0.004	&	0.055	 $\pm$	0.002	&	8.79	$\pm$	1	&	4.78	$\pm$	0.55	&	1, 1, 25	\\
070125	&	1.5477	&	84.09	$\pm$	8.41	&	935	$\pm$	165.6	&	3.73	$\pm$	0.52	&	O	&	0.087	$\pm$	0.005	&	0.052	 $\pm$	0.002	&	31.92	$\pm$	3.19	&	11.3	$\pm$	1.13	&	1, 26, 27	\\
070208	&	1.17	&	0.37			&	143.2	$\pm$	71.6	&	0.11	$\pm$	0.05	&	X	&	0.049	$\pm$	0.008	&	0.087	 $\pm$	0.01	&	0.04			&	0.14			&	2, 20, 2	\\
070306	&	1.5	&	6	$\pm$	5	&	300	$\pm$	97.5	&	1.33	$\pm$	0.86	&	X	&	0.083	$\pm$	0.02	&	0.078	$\pm$	 0.013	&	2.06	$\pm$	1.72	&	1.82	$\pm$	1.52	&	13, 20, 2	\\
070318	&	0.84	&	0.9	$\pm$	0.2	&	360.6	$\pm$	143.5	&	3.57	$\pm$	0.63	&	X	&	0.171	$\pm$	0.011	&	0.173	 $\pm$	0.008	&	1.31	$\pm$	0.29	&	1.35	$\pm$	0.3	&	13, 20, 2	\\
070411	&	2.95	&	10	$\pm$	8	&	474	$\pm$	154.1	&	0.24	$\pm$	0.11	&	X	&	0.034	$\pm$	0.006	&	0.04	$\pm$	 0.005	&	0.59	$\pm$	0.48	&	0.79	$\pm$	0.63	&	13, 13, 2	\\
070419A	&	0.97	&	0.24	$\pm$	0.1	&	53.2	$\pm$	31.5	&	0.02			&	O	&	0.027	$\pm$	0.002	&	0.062	$\pm$	 0.003	&	0.01	$\pm$	0	&	0.05	$\pm$	0.02	&	18, 16 , 3	\\
070508	&	0.82	&	8	$\pm$	2	&	378.6	$\pm$	138.3	&	0.58	$\pm$	0.93	&	X	&	0.066	$\pm$	0.04	&	0.064	 $\pm$	0.026	&	1.74	$\pm$	0.44	&	1.63	$\pm$	0.41	&	13, 13, 2	\\
070611	&	2.04	&	0.92			&	188	$\pm$	49	&	1.13	$\pm$	0.39	&	X	&	0.092	$\pm$	0.012	&	0.114	$\pm$	 0.01	&	0.39			&	0.6			&	2, 28, 2	\\
070721B	&	3.626	&	36.5			&	1715.3			&	0.11	$\pm$	0.01	&	X	&	0.021	$\pm$	0.001	&	0.023	$\pm$	 0.001	&	0.78			&	0.95			&	2, 29, 2	\\
070810	&	2.17	&	1.73			&	130	$\pm$	13	&	0.09	$\pm$	0.04	&	X	&	0.032	$\pm$	0.005	&	0.051	$\pm$	 0.006	&	0.09			&	0.23			&	2, 28, 2	\\
071003	&	1.1	&	38.5	$\pm$	1.8	&	2086	$\pm$	188	&	0.41	$\pm$	0.07	&	X	&	0.045	$\pm$	0.003	&	0.038	$\pm$	 0.002	&	3.92	$\pm$	0.18	&	2.79	$\pm$	0.13	&	5, 5, 2	\\
071010A	&	0.98	&	0.13			&	73	$\pm$	97.7	&	0.81	$\pm$	0.22	&	X	&	0.121	$\pm$	0.012	&	0.19	$\pm$	 0.013	&	0.1			&	0.23			&	13, 13, 2	\\
071010B	&	0.947	&	2.32	$\pm$	0.4	&	88	$\pm$	21	&	3.44	$\pm$	0.39	&	O	&	0.146	$\pm$	0.006	&	0.133	$\pm$	 0.004	&	2.49	$\pm$	0.43	&	2.06	$\pm$	0.36	&	1, 1, 2	\\
080210	&	2.64	&	5.13	$\pm$	2.13	&	329.4	$\pm$	132.8	&	0.14	$\pm$	0.06	&	X	&	0.031	$\pm$	0.005	&	 0.042	$\pm$	0.005	&	0.25	$\pm$	0.1	&	0.45	$\pm$	0.19	&	30, 16, 30	\\
080310	&	2.43	&	5.89	$\pm$	1.08	&	75.4	$\pm$	72	&	0.34	$\pm$	0.04	&	O	&	0.044	$\pm$	0.002	&	0.051	 $\pm$	0.001	&	0.57	$\pm$	0.11	&	0.78	$\pm$	0.14	&	16, 16, 3	\\
080319B	&	0.937	&	117.87			&	1261	$\pm$	65	&	0.03	$\pm$	0.01	&	O	&	0.015	$\pm$	0.002	&	0.015	$\pm$	 0.001	&	1.36			&	1.38			&	1, 1, 2	\\
080330A	&	1.5115	&	0.21	$\pm$	0.05	&	71	$\pm$		&	1			&	O	&	0.113			&	0.167			&	0.13	 $\pm$	0.03	&	0.29	$\pm$	0.07	&	31, 31, 31	\\
080413A	&	2.433	&	7.83	$\pm$	3.55	&	583.6	$\pm$	274.6	&	0.01	$\pm$	0	&	O	&	0.012			&	0.021	$\pm$	 0.001	&	0.06	$\pm$	0.03	&	0.17	$\pm$	0.08	&	3, 3, 3	\\
080413B	&	1.1	&	1.61	$\pm$	0.27	&	163	$\pm$	47.5	&	3.85	$\pm$	0.13	&	X	&	0.155	$\pm$	0.002	&	0.147	 $\pm$	0.001	&	1.95	$\pm$	0.32	&	1.75	$\pm$	0.29	&	1, 1, 32	\\
080603A	&	1.688	&	2.2	$\pm$	0.8	&	160	$\pm$	920	&	1.16	$\pm$	0.46	&	O	&	0.087	$\pm$	0.013	&	0.095	$\pm$	 0.009	&	0.83	$\pm$	0.3	&	0.99	$\pm$	0.36	&	13, 33, 2	\\
080710	&	0.85	&	1.68	$\pm$	0.22	&	200			&	0.23	$\pm$	0.02	&	O	&	0.057	$\pm$	0.001	&	0.075	$\pm$	 0.001	&	0.27	$\pm$	0.04	&	0.47	$\pm$	0.06	&	34, 13, 3	\\
080810	&	3.35	&	45	$\pm$	5	&	1470	$\pm$	180	&	0.11	$\pm$	0.02	&	X	&	0.02	$\pm$	0.002	&	0.022	$\pm$	 0.001	&	0.93	$\pm$	0.1	&	1.07	$\pm$	0.12	&	5, 5, 30	\\
080928	&	1.692	&	2.82	$\pm$	1.17	&	199.4	$\pm$	69.7	&	0.14	$\pm$	0.04	&	X	&	0.038	$\pm$	0.004	&	 0.053	$\pm$	0.003	&	0.21	$\pm$	0.09	&	0.39	$\pm$	0.16	&	30, 16, 30	\\
081007A    	&	0.5295	&	0.17			&	61.2	$\pm$	15.3	&	11.57			&	X	&	0.35			&	0.368			&	 1.05			&	1.16			&	35, 35, 36	\\
081008A	&	1.967	&	6.92	$\pm$	1.67	&	261.7	$\pm$	489.9	&	0.21	$\pm$	0.07	&	X	&	0.038	$\pm$	0.005	&	 0.045	$\pm$	0.004	&	0.5	$\pm$	0.12	&	0.71	$\pm$	0.17	&	30, 16, 30	\\
081203A	&	2.1	&	36.13	$\pm$	18.42	&	1791.8	$\pm$	899	&	0.12	$\pm$	0.02	&	O	&	0.025	$\pm$	0.002	&	0.026	 $\pm$	0.001	&	1.09	$\pm$	0.56	&	1.19	$\pm$	0.61	&	3, 3, 3	\\
090313A	&	3.375	&	3.2			&	240.1	$\pm$	223.5	&	1.04			&	X	&	0.066			&	0.075			&	0.7			 &	0.89			&	13, 16, 37	\\
090323	&	3.568	&	410	$\pm$	50	&	1901	$\pm$	343	&	17.8	$\pm$	19.6	&	X	&	0.103	$\pm$	0.043	&	0.045	$\pm$	 0.012	&	218.26	$\pm$	26.62	&	40.7	$\pm$	4.96	&	13, 13, 2	\\
090328	&	0.7354	&	13	$\pm$	3	&	1028	$\pm$	312	&	6.4	$\pm$	12	&	X	&	0.156	$\pm$	0.109	&	0.104	$\pm$	0.049	 &	15.74	$\pm$	3.63	&	7.05	$\pm$	1.63	&	13, 13, 2	\\
090417B	&	0.345	&	0.63			&	361.8			&	5.21	$\pm$	3.24	&	X	&	0.231	$\pm$	0.054	&	0.225	$\pm$	 0.035	&	1.69			&	1.59			&	38, 39, 38	\\
090423	&	8.23	&	11	$\pm$	3	&	491	$\pm$	200	&	14.6	$\pm$	2.7	&	X	&	0.116	$\pm$	0.008	&	0.088	$\pm$	0.004	 &	7.36	$\pm$	2.01	&	4.25	$\pm$	1.16	&	13, 13, 40	\\
090424A	&	0.544	&	4.6	$\pm$	0.9	&	273	$\pm$	50	&	11.57			&	X	&	0.231			&	0.161			&	12.29	$\pm$	 2.4	&	5.98	$\pm$	1.17	&	13, 13, 41	\\
090426	&	2.609	&	0.5	$\pm$	0.1	&	176.8	$\pm$	90.2	&	0.4	$\pm$	0.02	&	X	&	0.063	$\pm$	0.001	&	0.098	$\pm$	 0.001	&	0.1	$\pm$	0.02	&	0.24	$\pm$	0.05	&	42, 43, 44	\\
090618A	&	0.54	&	20			&	286.4	$\pm$	12.3	&	0.08	$\pm$	0.01	&	X	&	0.03	$\pm$	0.002	&	0.032	$\pm$	 0.001	&	0.9			&	1.04			&	45, 46, 47	\\
090902B	&	1.8829	&	1.77	$\pm$	0.01	&	596.8	$\pm$	17.3	&	6.2	$\pm$	2.4	&	O/X	&	0.163	$\pm$	0.024	&	0.15	 $\pm$	0.014	&	2.35	$\pm$	0.01	&	1.99	$\pm$	0.01	&	13, 13, 2	\\
090926A	&	2.1062	&	210	$\pm$	5.3	&	1016	$\pm$	25	&	4.06	$\pm$	0.81	&	O	&	0.074	$\pm$	0.006	&	0.04	$\pm$	 0.002	&	58.27	$\pm$	1.47	&	16.87	$\pm$	0.43	&	5, 5, 48	\\
091018	&	0.97	&	0.8	$\pm$	0.09	&	55	$\pm$	26	&	0.37	$\pm$	0.02	&	O	&	0.072	$\pm$	0.001	&	0.1	$\pm$	 0.001	&	0.21	$\pm$	0.02	&	0.4	$\pm$	0.04	&	5, 5, 49	\\
091029A	&	2.752	&	7.4	$\pm$	0.74	&	230	$\pm$	66	&	0.03	$\pm$	0	&	O	&	0.017	$\pm$	0	&	0.026			&	 0.11	$\pm$	0.01	&	0.26	$\pm$	0.03	&	5, 5, 32	\\
091127A	&	0.49	&	1.63	$\pm$	0.02	&	53.6	$\pm$	3	&	0.37	$\pm$	0.1	&	X	&	0.073	$\pm$	0.008	&	0.089	 $\pm$	0.006	&	0.44	$\pm$	0.01	&	0.65	$\pm$	0.01	&	5, 5, 50	\\
091208B	&	1.063	&	2.01	$\pm$	0.07	&	297.5	$\pm$	37.1	&	3.59			&	X	&	0.148			&	0.138			&	 2.21	$\pm$	0.08	&	1.9	$\pm$	0.07	&	5, 5, 51	\\
100219A	&	4.8	&	3.93			&	812			&	0.02	$\pm$	0	&	O	&	0.013	$\pm$	0.001	&	0.025	$\pm$	0.001	&	 0.04			&	0.12			&	34, 3, 3	\\
100418	&	0.6239	&	0.1	$\pm$	0.06	&	47.1	$\pm$	3.2	&	4.2	$\pm$	1.1	&	X	&	0.251	$\pm$	0.025	&	0.323	$\pm$	 0.021	&	0.31	$\pm$	0.2	&	0.52	$\pm$	0.33	&	5, 5, 52	\\
100814A	&	1.44	&	7.59	$\pm$	0.58	&	312	$\pm$	32	&	2	$\pm$	0.07	&	X	&	0.095	$\pm$	0.001	&	0.082	$\pm$	 0.001	&	3.41	$\pm$	0.26	&	2.54	$\pm$	0.19	&	5, 5, 53	\\
100901A	&	1.408	&	6.3			&	230			&	11.57			&	X	&	0.188			&	0.133			&	11.15			&	 5.6			&	13, 13, 2	\\
100906A	&	1.727	&	28.9	$\pm$	0.3	&	289	$\pm$	48	&	0.15	$\pm$	0.02	&	X	&	0.029	$\pm$	0.001	&	0.03	$\pm$	 0.001	&	1.24	$\pm$	0.01	&	1.3	$\pm$	0.01	&	5, 5, 30	\\
110205A	&	2.22	&	48.38	$\pm$	6.38	&	757	$\pm$	305	&	1.2	$\pm$	0	&	O/X	&	0.056			&	0.042			&	7.56	 $\pm$	1	&	4.32	$\pm$	0.57	&	1, 1, 2	\\
110801A	&	1.858	&	10.9	$\pm$	2.72	&	400	$\pm$	171	&	0.18	$\pm$	0.1	&	X	&	0.035	$\pm$	0.007	&	0.04	$\pm$	 0.006	&	0.66	$\pm$	0.17	&	0.85	$\pm$	0.21	&	1, 1, 30	\\
111209A	&	0.677	&	5.14	$\pm$	0.62	&	520	$\pm$	89	&	9.12	$\pm$	0.47	&	O	&	0.202	$\pm$	0.004	&	0.145	 $\pm$	0.002	&	10.5	$\pm$	1.27	&	5.38	$\pm$	0.65	&	1, 1, 54	\\
120119A	&	1.728	&	27.2	$\pm$	3.63	&	496	$\pm$	50	&	0.13	$\pm$	0.02	&	X	&	0.028	$\pm$	0.002	&	0.029	 $\pm$	0.001	&	1.05	$\pm$	0.14	&	1.16	$\pm$	0.15	&	1, 1, 2	\\
120326A	&	1.798	&	3.27	$\pm$	0.27	&	152	$\pm$	14	&	2.91	$\pm$	0.12	&	O	&	0.115	$\pm$	0.002	&	0.107	 $\pm$	0.001	&	2.16	$\pm$	0.18	&	1.88	$\pm$	0.16	&	1, 1, 55	\\
120404A	&	2.876	&	9			&	1043.8			&	0.06	$\pm$	0	&	O	&	0.021	$\pm$	0	&	0.029			&	0.2			 &	0.38			&	2, 56, 2	\\
120729A	&	0.8	&	1.24	$\pm$	0.27	&	559.8	$\pm$	36	&	0.06	$\pm$	0.01	&	O	&	0.037	$\pm$	0.001	&	0.059	 $\pm$	0.001	&	0.08	$\pm$	0.02	&	0.22	$\pm$	0.05	&	3, 3, 3	\\
121027A	&	1.773	&	1.58	$\pm$	0.08	&	130.3	$\pm$	83.2	&	0.14			&	X	&	0.04			&	0.06			&	 0.13	$\pm$	0.01	&	0.29	$\pm$	0.01	&	57, 58, 57	\\
121211A	&	1.023	&	0.63	$\pm$	0.54	&	194.2	$\pm$	26.3	&	0.44	$\pm$	0.28	&	X	&	0.078	$\pm$	0.019	&	 0.109	$\pm$	0.018	&	0.19	$\pm$	0.16	&	0.38	$\pm$	0.32	&	30, 59, 30	\\
130427A	&	0.3399	&	77.01	$\pm$	7.88	&	1250	$\pm$	150	&	0.43	$\pm$	0.05	&	O	&	0.05	$\pm$	0.002	&	0.036	 $\pm$	0.001	&	9.62	$\pm$	0.98	&	5.08	$\pm$	0.52	&	1, 1, 60	\\
130427B	&	2.78	&	3.16	$\pm$	1.75	&	354.6	$\pm$	111.9	&	0.21	$\pm$	0.09	&	X	&	0.038	$\pm$	0.006	&	 0.052	$\pm$	0.005	&	0.23	$\pm$	0.13	&	0.43	$\pm$	0.24	&	30, 61, 30	\\
130606A	&	5.91	&	28.3	$\pm$	5.2	&	2032	$\pm$	622	&	0.17	$\pm$	0.05	&	X	&	0.022	$\pm$	0.002	&	0.025	 $\pm$	0.002	&	0.67	$\pm$	0.12	&	0.86	$\pm$	0.16	&	30, 5, 30	\\
130907A	&	1.238	&	300			&	866	$\pm$	36	&	0.25	$\pm$	0.02	&	R	&	0.028	$\pm$	0.001	&	0.02			&	 12.03			&	5.89			&	62, 5, 62	\\
131030A	&	1.293	&	32.7	$\pm$	1.3	&	449	$\pm$	14	&	2.91	$\pm$	0.56	&	O	&	0.093	$\pm$	0.007	&	0.063	$\pm$	 0.003	&	14.1	$\pm$	0.56	&	6.55	$\pm$	0.26	&	5, 5, 63	\\
140311A	&	4.954	&	10			&	1201.5			&	1.3	$\pm$	1.1	&	R	&	0.056	$\pm$	0.018	&	0.055	$\pm$	0.012	&	 1.55			&	1.5			&	62, 64, 62	\\
140512A	&	0.725	&	7.25	$\pm$	0.61	&	826	$\pm$	202	&	0.21	$\pm$	0.02	&	X	&	0.047	$\pm$	0.001	&	0.052	 $\pm$	0.001	&	0.79	$\pm$	0.07	&	0.96	$\pm$	0.08	&	30, 5, 30	\\
140629A	&	2.276	&	4.4			&	251	$\pm$	55	&	0.42	$\pm$	0.09	&	O	&	0.05	$\pm$	0.004	&	0.059	$\pm$	0.003	 &	0.56			&	0.76			&	65, 65, 65	\\
141121A	&	1.469	&	8			&	209.4	$\pm$	67.4	&	3.8	$\pm$	0.5	&	R	&	0.119	$\pm$	0.006	&	0.095	$\pm$	0.003	 &	5.68			&	3.57			&	62, 66, 62	\\
151027A	&	0.81	&	3.3	$\pm$	0.41	&	313	$\pm$	105	&	0.69	$\pm$	0.21	&	X	&	0.079	$\pm$	0.009	&	0.083	$\pm$	 0.006	&	1.03	$\pm$	0.13	&	1.15	$\pm$	0.14	&	5 , 5, 67	\\
160131A	&	0.97	&	87	$\pm$	6.6	&	1284	$\pm$	454	&	0.13	$\pm$	0.02	&	X	&	0.027	$\pm$	0.001	&	0.024	$\pm$	 0.001	&	3.24	$\pm$	0.25	&	2.46	$\pm$	0.19	&	5, 5, 67	\\
160227A	&	2.38	&	5.56	$\pm$	0.36	&	222.4	$\pm$	55.4	&	1	$\pm$	0.06	&	X	&	0.067	$\pm$	0.002	&	0.069	 $\pm$	0.001	&	1.26	$\pm$	0.08	&	1.31	$\pm$	0.08	&	5 , 5, 67	\\
160509A	&	1.17	&	86			&	625	$\pm$	63	&	3.7	$\pm$	0.8	&	R	&	0.092	$\pm$	0.007	&	0.054	$\pm$	0.003	&	 36.4			&	12.33			&	62, 5, 62	\\
160625B	&	1.406	&	300			&	1374	$\pm$	29	&	22	$\pm$	4	&	R	&	0.148	$\pm$	0.01	&	0.06	$\pm$	0.003	&	 327.38			&	53.33			&	62, 5, 62	\\
161017A	&	2.013	&	6.87	$\pm$	0.72	&	699	$\pm$	113.1	&	0.57	$\pm$	1.61	&	X	&	0.055	$\pm$	0.059	&	0.058	 $\pm$	0.041	&	1.05	$\pm$	0.11	&	1.16	$\pm$	0.12	&	5, 68, 67	\\
170405A	&	3.51	&	9.01			&	1602	$\pm$	38.8	&	0.06	$\pm$	0.01	&	X	&	0.02	$\pm$	0.001	&	0.028	 $\pm$	0.001	&	0.18			&	0.36			&	67, 69, 67	\\
171010A	&	0.33	&	22			&	227.2	$\pm$	9.3	&	6.48	$\pm$	27.09	&	O	&	0.162	$\pm$	0.254	&	0.098	$\pm$	 0.102	&	28.76			&	10.54			&	3, 5, 3	\\
180115A	&	2.487	&	3.64	$\pm$	0.53	&	116.9			&	0.2	$\pm$	0.04	&	X	&	0.038	$\pm$	0.003	&	0.05	$\pm$	 0.003	&	0.26	$\pm$	0.04	&	0.46	$\pm$	0.07	&	67, 70, 67	\\
180620B	&	1.1175	&	3.04	$\pm$	0.03	&	372	$\pm$	105	&	2.8	$\pm$	0.44	&	X	&	0.127	$\pm$	0.007	&	0.116	$\pm$	 0.005	&	2.45	$\pm$	0.02	&	2.04	$\pm$	0.02	&	5, 71, 67	\\
180720B	&	0.654	&	33.97	$\pm$	0.01	&	1052	$\pm$	26	&	1.08	$\pm$	0.27	&	X	&	0.072	$\pm$	0.007	&	0.053	 $\pm$	0.003	&	8.83			&	4.8			&	5, 72, 67	\\
180728A	&	0.117	&	0.23	$\pm$	0.01	&	108	$\pm$	8	&	1.82	$\pm$	1.08	&	X	&	0.189	$\pm$	0.042	&	0.232	 $\pm$	0.035	&	0.42	$\pm$	0.02	&	0.63	$\pm$	0.03	&	5, 73, 67	\\
181020A	&	2.938	&	82.8	$\pm$	11.6	&	1461	$\pm$	225	&	0.13	$\pm$	0.04	&	X	&	0.021	$\pm$	0.003	&	0.02	 $\pm$	0.002	&	1.82	$\pm$	0.26	&	1.68	$\pm$	0.23	&	5, 74, 67	\\
190114C	&	0.42	&	27.03	$\pm$	0.24	&	929.3	$\pm$	9.4	&	0.1	$\pm$	0.04	&	X	&	0.032	$\pm$	0.005	&	0.032	 $\pm$	0.003	&	1.38	$\pm$	0.01	&	1.39	$\pm$	0.01	&	5, 5, 67	\\
\enddata
\tablenotetext{a}{The jet-break is identified in radio (R), optical (O), or X-ray (X), respectively.}
\tablenotetext{b}{References for $E_{\gamma,iso}$, $E_{p,i}$ and $T_{jet}$, respectively.}
\tablerefs{(1) Demianski et al. 2017; (2) Lu et al. 2012; (3) Wang et al. 2018a; (4) Amati et al.2008; (5) Minaev \& Pozanenko 2020;
(6) Bloom et al. 2003; (7) Bj{\"o}rnsson et al. 2002;
(8) Greiner et al. 2003; (9) Pandey  et al. 2004;
(10) Maiorano et al. 2006; (11) Jakobsson et al. 2004; (12) Vaughan et al. 2006; (13) Song et al. 2018;
(14) Kamble et al. 2009; (15) de Ugarte Postigo et al. 2007; (16) Kann et al. 2010; (17) Perri et al. 2007;
(18) Wang et al. 2018b; (19) $https://gcn.gsfc.nasa.gov/notices\b{ }s/148522/BA/$;
(20) Butler et al. 2007; (21) Burrows et al. 2006; (22) Liu et al. 2008; (23) Krimm et al. 2007;
(24) Ruiz-Velasco et al. 2007; (25) Gomboc et al. 2008; (26) Bellm et al. 2008; (27) Updike et al. 2008;
(28) Rossi et al. 2008; (29) $https://gcn.gsfc.nasa.gov/notices\b{ }s/285654/BA/$; (30) Xi et al. 2017; (31) Guidorzi et al. 2009;
(32) Filgas et al. 2011; (33) Guidorzi et al. 2011; (34) Ruffini et al. 2016;
(35) Ghirlanda et al. 2010; (36) Jin et al. 2013; (37) Melandri et al. 2010; (38) Holland et al. 2010;
(39) $https://gcn.gsfc.nasa.gov/notices\b{ }s/349450/BA/$;
(40) Laskar et al. 2014; (41) Jin et al. 2013;
(42) Th{\"o}ne et al. 2011;
(43) Antonelli et al. 2009; (44) Nicuesa Guelbenzu et al. 2011; (45) Campana et al. 2011; (46) Zhang et al. 2012; (47) Page et al. 2011;
(48) Swenson et al. 2010; (49) Wiersema et al. 2012; (50) Troja et al. 2012;
(51) Uehara et al. 2012; (52) de Ugarte Postigo et al. 2018; (53) Nardini et al. 2014; (54) Kann et al. 2018;
(55) Urata et al. 2014; (56) Ukwatta et al. 2012; (57) Wu et al. 2013; (58) Peng et al. 2013;
(59) Heussaff et al. 2013; (60) Maselli et al. 2014; (61) Barthelmy et al. 2013;
(62) Kangas \& Fruchter 2019; (63) Huang et al. 2017; (64) Krimm et al. 2014a; (65) Hu et al. 2019;
(66) Krimm et al. 2014b; (67) This paper; (68) Zhou et al. 2020; (69) Hui et al. 2017; (70) $https://gcn.gsfc.nasa.gov/notices\b{ }s/805318/BA/$;
(71) $https://gcn.gsfc.nasa.gov/gcn3/22825.gcn3$;
(72) Cherry et al. 2018; (73) $https://gcn.gsfc.nasa.gov/gcn3/23053.gcn3$; (74) Veres et al. 2018.  }
\end{deluxetable}

\clearpage
\begin{deluxetable}{ccccccccccccccccccccccccc}
\tabletypesize{\scriptsize} \tablecaption{Results of the linear
regression analysis for GRBs with jet breaks. $R$ is the Spearman
correlation coefficient, and $P$ is the chance probability} \tablewidth{0pt}
%\tabletypesize{\tiny}

\tablehead{ \colhead{Correlations}& \colhead{Expressions}&
\colhead{$R$}& \colhead{$P$} }

\startdata
\hline
$E_{\rm \gamma,iso}-E_{\rm p,i}$ & $\log E_{\rm \gamma,iso}=(49.61\pm 0.09)+(1.27\pm0.04)\times \log E_{\rm p,i}$ & 0.74 & $4.83\times10^{-24}$  \\
$E_{\rm jet}(ISM)-E_{\rm p,i}$ & $\log E_{\rm jet}(ISM)=(48.17\pm 0.09)+(0.77\pm0.03)\times \log E_{\rm p,i}$ & 0.41 & $6.12\times10^{-7}$ \\
$E_{\rm jet}(Wind)-E_{\rm p,i}$ & $\log E_{\rm jet}(Wind)=(48.84\pm 0.05)+(0.51\pm0.02)\times \log E_{\rm p,i}$ & 0.41 & $9.72\times10^{-7}$ \\
$E_{\rm \gamma,iso}-\theta_{\rm jet}(ISM)$ & $\log E_{\rm \gamma,iso}=(51.84\pm 0.03)+(-0.85\pm0.03)\times \log \theta_{\rm jet}(ISM)$ & -0.36 & $1.29\times10^{-5}$ \\
$E_{\rm \gamma,iso}-\theta_{\rm jet}(Wind)$ & $\log E_{\rm \gamma,iso}=(50.68\pm 0.04)+(-1.82\pm0.03)\times \log \theta_{\rm jet}(Wind)$ & -0.70 & $2.31\times10^{-21}$ \\
\hline
$E_{\rm \gamma,iso}(T_{\rm j,z},E_{\rm p,i})\tablenotemark{a}$ & $\log E_{\rm \gamma,iso}=(49.89\pm 0.46)+(0.12\pm0.02)\times \log T_{\rm j,z}$ & 0.80 & $9.76\times10^{-12}$ \\& $+(0.99\pm0.17)\times \log E_{\rm p,i}$ &  &  & \\
$E_{\rm \gamma,iso}(T_{\rm j,z},E_{\rm p,i})\tablenotemark{b}$ & $\log E_{\rm \gamma,iso}=(50.22\pm 0.12)-(0.14\pm0.01)\times \log T_{\rm j,z}$ & 0.84 & $1.82\times10^{-22}$ \\& $+(1.23\pm0.04)\times \log E_{\rm p,i}$ &  &  & \\
$E_{\rm \gamma,iso}(T_{\rm j,z},E_{\rm p,i})\tablenotemark{c}$ & $\log E_{\rm \gamma,iso}=(50.05\pm 0.26)-(0.02\pm0.01)\times \log T_{\rm j,z}$ & 0.80 & $5.22\times10^{-32}$ \\& $+(1.13\pm0.09)\times \log E_{\rm p,i}$ &  &  & \\
\hline
$E_{\rm jet}(ISM)(T_{\rm j,z},E_{\rm p,i})$ & $\log E_{\rm jet}(ISM)=(45.08\pm 0.21)+(0.67\pm0.02)\times \log T_{\rm j,z}$ & 0.90 & $4.70\times10^{-50}$ \\& $+(0.84\pm0.07)\times \log E_{\rm p,i}$ &  &  & \\
$E_{\rm jet}(Wind)(T_{\rm j,z},E_{\rm p,i})$ & $\log E_{\rm jet}(Wind)=(46.77\pm 0.14)+(0.45\pm0.01)\times \log T_{\rm j,z}$ & 0.90 & $5.62\times10^{-50}$ \\& $+(0.56\pm0.05)\times \log E_{\rm p,i}$ &  &  & \\
\enddata
\tablenotetext{a}{The optical jet break sample.}
\tablenotetext{b}{The X-ray jet break sample.}
\tablenotetext{b}{The total jet break sample.}
\end{deluxetable}

\end{document}